\documentclass{aa}  

\usepackage{graphicx}
\usepackage{txfonts}
\usepackage{hyperref}

\newcommand{\kms}{\hbox{km $\cdot$ s$^{-1}$}}
\newcommand{\logg}{\hbox{log\,$\it g$}}
\newcommand{\feh}{\hbox{$\rm [M/H]$}}
\newcommand{\teff}{\hbox{$T_{\rm eff}$}}
\newcommand{\vsini}{\hbox{$v_{\sin i}$}}

\begin{document}

\title{A Spectroscopic Modelling Method for the Detached Eclipsing Binaries to Derive Atmospheric Parameters 
}%

\titlerunning{A Method to Derive Atmospheric Parameters of Binaries}

\author{
Xiang-Lei Chen\inst{1,2}
\and
A-Li Luo\inst{1,2,3,4}\thanks{A-Li Luo  ~~  email: lal@nao.cas.cn}
%\corref{mycorrespondingauthor}
%\cortext{[mycorrespondingauthor]{Corresponding Author}
\and
Jian-Jun Chen\inst{1,2}\thanks{Jian-Jun Chen ~~  email: jjchen@nao.cas.cn}
\and
Rui Wang\inst{1}
\and
Xiao-Bin Zhang\inst{1,2}
\and
Wen Hou\inst{1}
\and
Bo Qiu\inst{4}
\and
Fang Zuo\inst{1}
}

\institute{
CAS Key Laboratory of Optical Astronomy, National Astronomical Observatories, Beijing 100101, China.\\
\and
School of Astronomy and Space Science, University of Chinese Academy of Sciences, Beijing 100049, China\\
\and
School of Information Management \& Institute for Astronomical Science, Dezhou University, Dezhou 253023, China\\
\and
School of Electronic Information Engineering, Hebei University of Technology, Tianjin 300401, China
}

\date{Received March 7, 2022; accepted March 00, 0000}

  \abstract
  {
	Based on luminosity contributions, we develop a spectroscopic modelling method to derive atmospheric parameters of component stars in binary systems. The method is designed for those spectra of binaries which show double-lined features due to the radial velocities differences between the component stars. We first derive the orbital parameters and the stellar radii by solving the light and radial velocity curves. Then the luminosity contributions in different phases can be calculated. The synthesised double-lined spectra model is constructed by superposing theoretical single-star spectra according to the luminosity contributions. Finally, we derive the atmospheric parameters of each component star by the model fitting method. For multi-epoch double-lined spectra observed by the Large sky Area Multi-Object Spectroscopic Telescope (LAMOST) Medium Resolution Survey ($R \sim 7500$), our method gives robust results for detached eclipsing binary systems observed in different orbital phases. Furthermore, this method can also be applied to other spectroscopic data with different resolutions as long as the systems are detached eclipsing binaries with nearly spherical stars.
  }
  \keywords{methods: data analysis -- (stars:) binaries: spectroscopic -- (stars:) binaries: eclipsing 
  }

\maketitle
%
%________________________________________________________________
\section{Introduction}
\label{sect:intro}

The binary stars play an important role in the study of stellar evolution. At formation time, stars are always in clusters. In their main sequence time, stars are fractionally in binary systems, depending on their spectral types. \cite{Ragh2010} considered the fraction is larger than 50\% for FGK stars. \cite{Dubinaryfraction2013} demonstrated that the stellar binarity could be 20\%  to 80\% for different spectral types. By analysing binary stars, stellar structure and evolution can be constrained (\cite{Han2020}). For example, \cite{Izzard2009} gave constraints on the mass of stars that have efficient third dredge up by comparing the observed Carbon-enhanced metal-poor (CEMP) stars with the binary population synthesis (BPS) study results. Another important objects are the Type Ia supernovae (SNe Ia). As classical cosmological distance tracers, the formation channels of SNe Ia affect the Hubble constant measurement and related cosmological assumptions and theories. Traditionally, the binary system consisting of a carbon-oxygen white dwarf and a main sequence/giant/helium donor star was thought to be a progenitor of an SNe Ia (\cite{Hoyle1960, Whelan1973}). During the later studies, \cite{Iben1984, Webbink1984, Han1998} proposed the double degenerate scenario that two carbon-oxygen white dwarfs merging may also cause the explosion of an SNe Ia. Besides the stellar evolution, the binary interaction can change the galaxy spectral energy distribution (SED). \cite{Han2007} found that the radiation of hot subdwarfs, which are the evolution results of binary interactions, cause the far-ultraviolet excess in the early-type galaxy spectra. \cite{Chen2015} proposed that the galaxy soft X-ray emissions come from the white dwarf accretion in binary systems. The study of binary stars will also contribute to our understanding of the early evolution of the Universe. The hypothesis about the re-ionising photons is that they come from massive single stars. But the photon number is several times less than the required number. Due to the high proportion of binarity of the massive stars, \cite{Gotberg2020} and \cite{Secunda2020} believed that the ionising photons produced by massive binary stars have an important impact on the re-ionisation process in the early Universe.

Depending on whether the mass transfer between the component stars has started, binary systems are divided into detached, semi-detached and contact binaries. This work concentrates on the detached binaries. Detailed analysis of the component stars' characteristics at this stage can not only help to understand the formation of binary stars but also limit the later evolution of mass transfer. Besides mass transfer, the inclination of the orbit is another factor affecting the shape of the light curve. If the inclination of the orbital plane is close to 90$^{\circ}$, a binary system is observed as an eclipsing binary star in time domain photometry. Orbital parameters of the eclipsing binaries can be calculated using both the light curve and the radial velocity curve data. And if the orbital period of the binary is shorter than five years, the system can be spectroscopically identified (\cite{apogeebinary}). The orbit of the component stars around each other causes the relative change of radial velocities. The binary spectra thus show double-peaked features and they are called `SB2' spectra. One can fit the profile of a binary spectrum by simply superposing two single-star spectra, but the atmospheric parameters of component stars may not be correctly derived because of spectra mixing. The spectra mixing affects the profile and strength of both continuum and spectral lines. In different orbital phases, the shifted continuum has varying degrees of impact on line depth. Therefore, modelling the spectra of binary stars needs to take into account not only the intrinsic luminosity contribution of each star but also the mixing effects in different phases.

The data product of the Medium Resolution Survey of the Large sky Area Multi-Object Spectroscopic Telescope (hereafter LAMOST MRS) includes both spectra and atmospheric parameters of stars. The resolving power of LAMOST MRS is $R=7500$, and their wavelength ranges are within 4950-5350 \AA \ in blue arms and 6300-6800 \AA \ in red arms, respectively. \footnote{http://dr7.lamost.org/v2.0/doc/mr-data-production-description}. The seventh data release (DR7) of LAMOST published more than 3.8 million MRS spectra including single exposure and combined spectra, and stellar parameters of 0.78 million stars with combined spectra. The MRS targeting strategies contain a series of time-domain fields (\cite{MRSLiuChao}) and result in a large number of multiple star system spectra, including SB1 (the spectra lines are single-peaked but the radial velocities vary among different epochs), SB2 and ST (triple-lined spectra of trinary stars) spectra. \cite{Lidoubleline} selected 3175 SB2 and 132 ST candidates using cross-correlation functions between LAMOST MRS spectra and theoretical models. \cite{Zhangdoubleline} applied convolution neural network to develop a distinguishing model and gave a catalogue of 2198 SB2 candidates. Both \cite{Lidoubleline} and \cite{Zhangdoubleline} suggested that, the spectra lines of the component stars in a binary system show clear double-peaked features when their RV differences are larger than 50 km/s with LAMOST MRS resolving power of R$\sim$7500.

\begin{figure}
	\center
	\includegraphics[width=\columnwidth, angle=0]{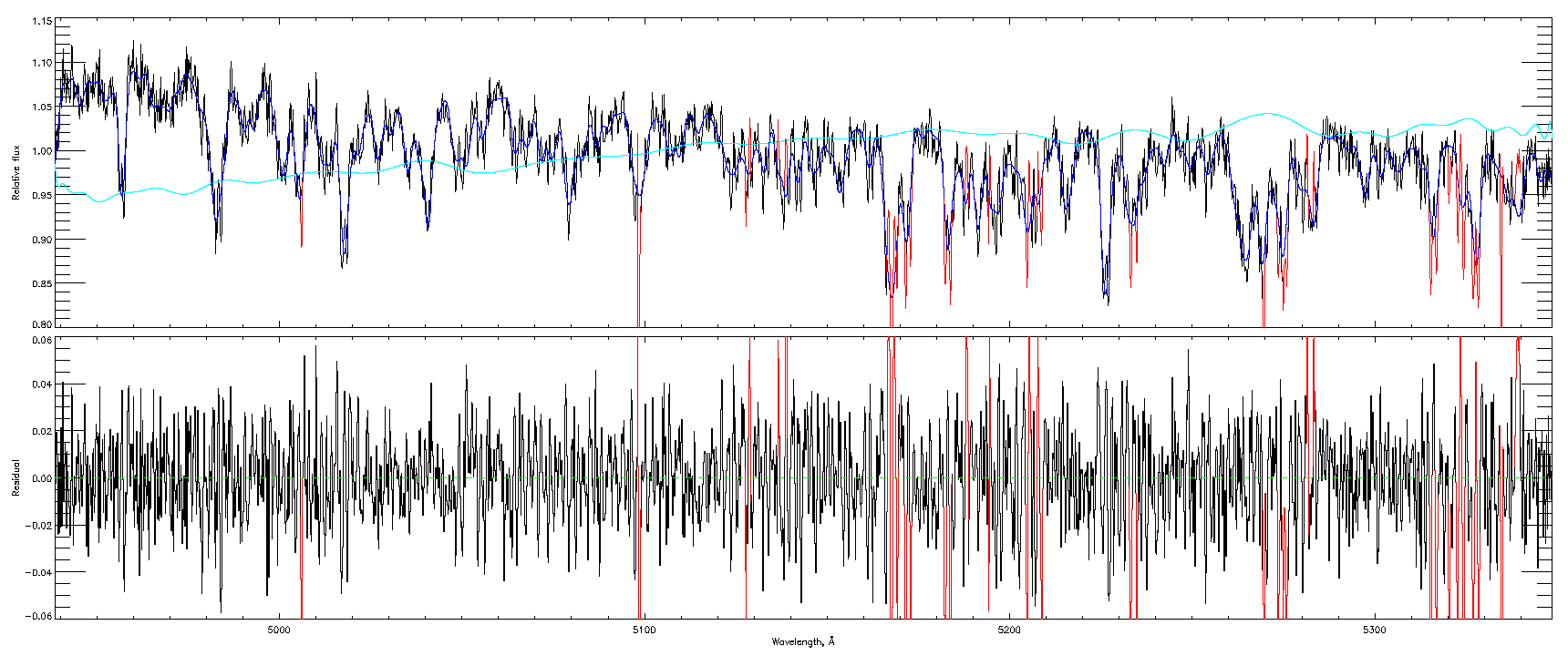}
	\caption{LASP fitting result of an SB2 spectrum. The upper panel is the spectrum, and the lower panel is the fitting residual. The black line in the upper panel is the pseudo-continuum normalised LAMOST MRS spectrum, the blue line is the best fitting spectrum, and the cyan line represents the polynomial correction between the observed spectrum and the model. Red lines in both panels represent the masked wavelength.}
	\label{lasp}
\end{figure}

It is necessary to build spectral model to derive atmospheric parameters of component stars for the LAMOST MRS SB2 spectra. The LAMOST Stellar Parameter pipeline (LASP) is the official data processing pipeline for LAMOST MRS spectra (\cite{LASPWuYue,DR1}). It was designed to derive atmospheric parameters for single stars based on the model fitting method, but was not suitable to be applied to SB2 spectra. Figure~\ref{lasp} shows an example of an SB2 spectrum and the "best-fit" spectrum by LASP. The upper panel is the best fitting result of the blue band, and the lower panel shows the residuals. The black line in the upper panel is the pseudo-continuum normalised observed spectrum, and the blue line is the best fitting model. The red lines stand for the LASP automatic masked points, which will not be considered in the fitting procedure. The cyan line represents the polynomial correction between the observed spectrum and the model. Mg I $\lambda$5185, for example, is a spectral line that shows clearly a double-peaked feature, but the peaks are masked. The best-fitting model has significantly broader lines than any of the component star spectrum. The derived parameters of the SB2 spectra can not reflect the real characteristics of the component stars, so the LAMOST parameter catalogue does not contain atmospheric parameters of SB2 spectra. It's worth developing a parameter deriving method for this kind of data. 

Therefore, we mainly focus on the LAMOST MRS SB2 spectra and propose a binary star spectral modelling method to derive the atmospheric parameters for the binary stars. The method synthesises the SB2 spectral model by combining theoretical spectra with the light curve solutions. The SB2 model is used to fit the observed binary spectra and to derive stellar parameters of the component stars. Furthermore, the method can be extended to spectroscopic data with a variety of resolutions, as long as the binary system has similar characteristics to our criteria (see details in Sect~\ref{sect:method}).

The remainder of this paper is organised as follows. Sect.~\ref{sect:method} introduces the SB2 spectra fitting method, as well as a brief data description. Sect.~\ref{sect:parameters} gives two examples of atmospheric parameters derivation. Sect.~\ref{sect:summary} is a summary section. Appendix~\ref{appendix-a} provides full procedures and formulae for the researchers. In Appendix~\ref{appendix-b} we discuss the method performance on the unresolved binary spectra observed in near-eclipsing phases.

\section{Method}
\label{sect:method}

\subsection{Data}
\label{subsect:data}

The sources in this work are observed by both LAMOST MRS and  Kepler/K2 (\cite{Kepler2010Bat, Kepler2010Cal, Kepler2010Koc, Kepler2011Bor}) or TESS (\cite{TESS2015Ric, TESS2017Lun, TESS2018Oel, TESS2018Sta}). And the targets must have at least three SB2 spectra in LAMOST MRS DR7 to evaluate our method's self-consistency.

\subsection{Double-lined Binary Spectral Model}
\label{subsect:sb2model}

\begin{figure}
	\center
	\includegraphics[width=\columnwidth, angle=0]{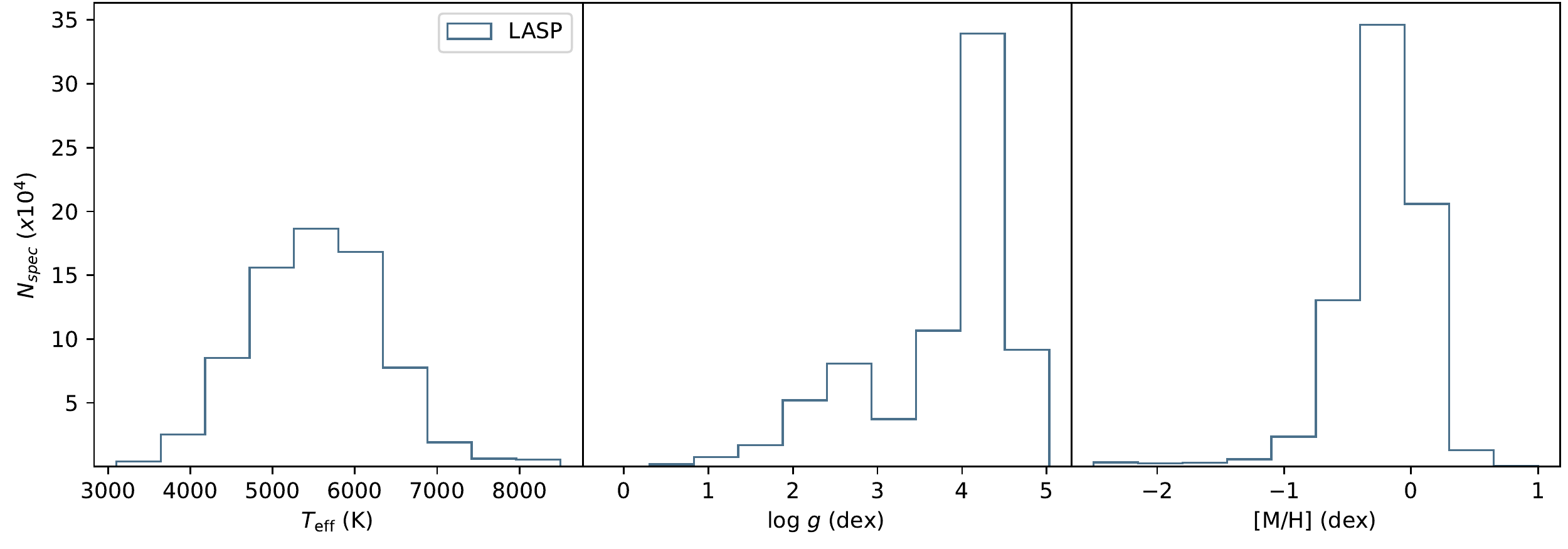}
	\caption{Stellar parameter distribution derived by the LASP. Using the data from \it{LAMOST MRS Stellar Parameter Catalogue} \rm DR7 v2.0.} 
	\label{lasp-param}
\end{figure}

A binary spectrum contains light from both component stars. The observed spectrum of each binary is affected by the distance to us, the stellar luminosity and the spectral line profiles. The line profiles are determined by the atmospheric parameters like effective temperature, the surface gravity, the metallicity, etc. Thus the spectroscopic model of the binary star can be explained as follows:

\begin{equation}
	%\begin{aligned}
		%F_{obs} = \frac{k}{4\pi}[\frac{(RV_1 \times L_1)A_1}{R_1^2}+\frac{(RV_2 \times L_2)A_2}{R_2^2}].
		F_{binary}(\lambda) = F_1(\lambda) + F_2(\lambda)
		\label{equ-base}
	%\end{aligned}
\end{equation}

\begin{equation}
	%\begin{aligned}
		%F_{obs} = \frac{k}{4\pi}[\frac{(RV_1 \times L_1)A_1}{R_1^2}+\frac{(RV_2 \times L_2)A_2}{R_2^2}].
		F_{i}(\lambda) = C \cdot L_i \cdot P_i(\lambda_i, T_{\rm eff,i}, \log \, \it g_i, \rm [M/H])
		\label{equ-base-param}
	%\end{aligned}
\end{equation}

$F_{binary}$ in Eq~\ref{equ-base} is the flux of the binary. $F_1$ and $F_2$ is the flux of the component stars, separately. $\lambda$ is the observed wavelength. $\lambda_i$ stands for the shifted wavelength of each component in a specific orbital phase. $C$ stands for the relationship between the luminosity and the observed flux. In the shorter wavelength range, $C$ is considered a constant and depends on the binary's distance to us. So $C$ has the same value for each component star. $L_i$ represents the luminosity of each star. $P_i$ means the spectral line profiles. In this work, we use \teff, \logg \space and \feh \space to generate the theoretical continuum and line profiles simultaneously. For a binary system, \teff \space and \logg \space are derived for each component while the same \feh \space is assigned. We omit the projected stellar rotation velocity \vsini \space to generate the line profiles. The degeneration in surface gravity, rotational velocity, line mixing in binary spectra, and line spread function (LSF) of telescopes in line broadening can affect the accurate measurement of \logg. See Sect~\ref{subsec:eg01} for more details. In a binary spectrum, the profiles of one component will be affected by the line mixing and the continuum of the other. The Doppler shifts of the continuum caused by the relative motion in different orbital phases affect the profiles differently.

For spectra without absolute flux calibration, the spectra are pseudo-continuum normalised before deriving stellar parameters, and only the line features are used. After the normalisation, $C$ in Equ~\ref{equ-base-param} will be ignored for one binary system. The binary star model can be explained as:

\begin{equation}
	%\begin{aligned}
		%F_{obs} = \frac{k}{4\pi}[\frac{(RV_1 \times L_1)A_1}{R_1^2}+\frac{(RV_2 \times L_2)A_2}{R_2^2}].
		f_{binary}(\lambda) = \frac{L_1}{L_1+L_2} \cdot P_1(\lambda_1, param) + \frac{L_2}{L_1+L_2} \cdot P_2(\lambda_2, param)
		\label{equ-normed}
	%\end{aligned}
\end{equation}

$f_{binary}$ is the normalised binary spectrum.  To obtain the luminosity contribution $\frac{L_i}{L_1+L_2} \ (i=1 \ \rm{or} \ 2)$ of each star, we choose to model the binary spectra of the detached eclipsing binaries, because the orbital parameters and luminosity contributions of the eclipsing binaries can be derived by combining the light curves and the radial velocity curves. In different phases, the component star in a detached system is not distorted significantly by the gravitation of its companion. When the stars are approximated as spheres, the projected areas of the component stars can be simply regarded as circular surfaces in the modelling. 

When the light curve solution is obtained, the binary spectral model with continuum is described as:

\begin{equation}
\begin{aligned}
		F_{binary}( \lambda ) &= A_1 \cdot L_1 \cdot P_1(\lambda_1, T_{\rm eff,1}, \log \, g_{\rm1}, \rm[M/H]) \\
		 &+ A_2 \cdot L_2 \cdot P_2(\lambda_2, r_{\rm T} \cdot T_{\rm eff,1}, \log \, g_{\rm 2}, \rm [M/H]),
\end{aligned}
\label{equ-area}
\end{equation}

and then this model spectra will be pseudo-normalised to derive stellar parameters. In this model, the luminosity contribution of each component is determined not only by its intrinsic brightness ($L_i $) but also by the projection area ($A_i$) in a specific phase. $A_1$ and $A_2$ are calculated using stellar radius $R_1$ and $R_2$. The radii are obtained by analysing the orbital motion of the binary system. The light curves used in this work are from Kepler/K2 or TESS mission, and the radial velocity curves are reconstructed using multi-epoch LAMOST MRS SB2 spectra. We use Wilson-Devinney binary star modelling code (WD, \cite{WD1971, WD1979, WD1990, WD2008, WD2014}) through a Python UI, PyWD2015 (\cite{PyWD}), to derive the orbital parameters. After analysing the light curve, we get orbital period $P$, stellar radius $R_1$ and $R_2$, semi-major axis $SMA$, orbital inclination $i$, eccentricity $e$, periastron longitude $\omega$, mass ratio $q$ and T$_{eff}$ ratio $r_{\rm T}$ for the next steps. The observation epoch of the spectrum is folded into phase angle $\theta$.

The theoretical model is computed using the SPECTRUM (\cite{SPECTRUM}) base on the Kurucz ATLAS9 (\cite{ATLASnine}), under the local thermodynamic equilibrium (LTE) and the plane-parallel atmosphere assumptions to generate the flux density spectra. Then we use \emph{the Payne} (\cite{Payne2019}) to interpolate the spectra of the component stars. \emph{The Payne} is a stellar label fitting method that consists of several ingredients. The `interpolator' of \emph{the Payne} applies neural networks to produce spectral model fluxes for a set of arbitrary labels. The default neural network architectures, which have two hidden layers and ten neurons for each layer, are adopted in this work. The training set of \emph{the Payne} is the Kurucz model mentioned above and the labels are \teff, \logg \space and \feh, and the parameter range of the training set is: $4500 \leq T_{\rm eff} \leq 8000 $ K, $2.0 \leq$ log\,$\it g \leq \rm 5.0 $ dex, $-2.0 \leq \rm [M/H] \leq 0.5$ dex. The ranges cover most of the LAMOST MRS observation data. Figure~\ref{lasp-param} shows the released stellar parameter distribution of LAMOST MRS DR7 v2.0. We ignore the spectral model with \teff \space lower than 4500 K because the mixing of the molecular bands will be more complicated. After the single star spectra are interpolated by \emph{the Payne}, we synthesise the SB2 spectra with different RVs for the component stars and the projection areas that are calculated applying the light curve solution. The pseudo-continuum normalised SB2 spectra are used to fit the observed spectra with the least square method. The SB2 model generation and the spectral fitting are iterative. In each iteration, we synthesise the SB2 model with a new set of parameters to fit the observed spectra. Both blue and red bands are generated and applied in the fitting. This is unlike the LASP, which uses only the blue band in the fitting procedure. Although the LAMOST blue band contains more information than its red band in parameter deriving (\cite{ZhangBoMRSinfo}), blue band is not sensitive enough to luminosity. \cite{MRSvalue2021Chen} suggested that the red band should also be used to derive atmospheric parameters, especially to derive \logg.

$\lambda_1$ and $\lambda_2$ in Equ~\ref{equ-area} are the shifted wavelength of the component stars caused by their radial velocities $RV_1$ and $RV_2$. The radial velocities are automatically determined in the fitting procedure. $T_{\rm eff1}$ is the effective temperature of the hotter star and the effective temperature of the cooler star is represented by $r_{\rm T} \cdot T_{\rm eff1}$. The initial value of $r_{\rm T}$ comes from the light curve solution, and we set it to be variable in the range of $\pm 0.1$, although the true ratio of \teff \space is constant.  The variation of $r_{\rm T}$ actually reflects the change of relative shift of two continua in the SB2 spectra. The application of $r_{\rm T}$ can limit the spectral interpretation by the light curve characteristics and reduce the iteration times. log\,$\it g_{\rm 1}$ and log\,$\it g_{\rm 2}$ are the surface gravity of the two stars, respectively. Only one metallicity is derived for each spectrum. 

In the non-eclipsing phases, the two component stars contribute all their fluxes to the binary spectrum. The projection area $A_i$ is calculated according to the area of the circle. While in the eclipsing phases, eclipsing depth and surface blocking should be considered in luminosity contribution. When judging the relative distance $d$ of the component stars, the eclipsing area can be calculated simultaneously. $d$ can be calculated using Equ~\ref{equ-app-d} in Appendix~\ref{appendix-a}. Assuming that $R_1>R_2$, the relative positions of the component stars is obtained by comparing the values of $d$, $R_1+R_2$, $R_1-R_2$:

a. $d>(R_1+R_2)$: The luminosity distribution in non-eclipsing phases is discussed before.

b. $d<(R_1-R_2)$: The total eclipsing phase.

c. $(R_1+R_2)>d>(R_1-R_2)$: During the partially eclipsing phases, the shielded areas are approximated to the area of a portion of a circle cut by a line segment. The area calculations are described in Appendix~\ref{appendix-a}. 

We do not consider the reflection effect and the limb darkening effect yet in this work, although they should be taken into consideration for the eclipsing phase spectra. Double-lined features in one spectrum indicate that the component stars have similar luminosity. Otherwise, the features of the less luminous star would be veiled by the other. Sometimes, the temperature difference between the primary and secondary star can be large for the situation where the cooler star has a larger radius than the hotter star. In the eclipsing phases, the reflection effect may change the magnitudes and thus the luminosity contributions. And in these phases, the limb darkening effect may reduce the luminosity contribution of the eclipsed star, which increases the difficulty of deriving its atmospheric parameters. For the LAMOST MRS spectra, when $\Delta RV$ between the two stars are smaller than 50 \kms, the stellar parameters derived by fitting the asymmetry of blended lines are quite different from those by the double-lined features (see Appendix~\ref{appendix-b}). This may be caused by the resolving power or the noise of the spectra. The parameters from the blended lines are excluded in the final results, and the neglect of the reflection effect and the limb darkening effect is ignored for LAMOST MRS SB2 spectra. But we note that for the data with higher resolution, the effects should be considered more carefully.

\subsection{Continuum Influence and Method Uncertainty}
\label{subsect:conti}

\begin{figure*}
	\center
	\includegraphics[width=16.0cm, angle=0]{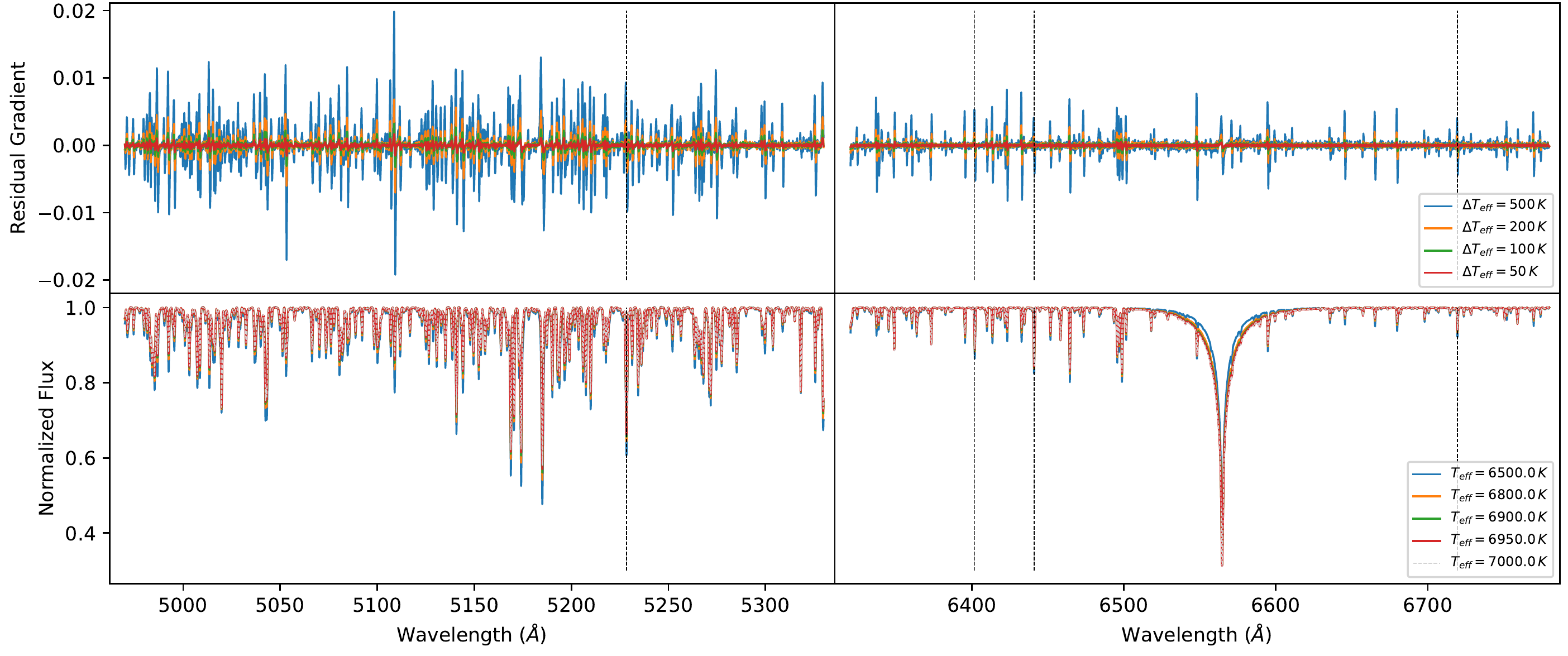}
	\caption{The upper panel shows the gradient of line intensity differences of four \teff \space gaps: $7000 - 50/100/200/500$ K. The lower panel contains the corresponding spectra. Different colours represent spectra with different \teff. The grey dashed lines indicate the spectral lines we chose to produce Figure~\ref{lineratio-teff} and Figure~\ref{lineratio-R}.}
	\label{teffgra}
\end{figure*}

\begin{figure}
	\center
	\includegraphics[width=\columnwidth, angle=0]{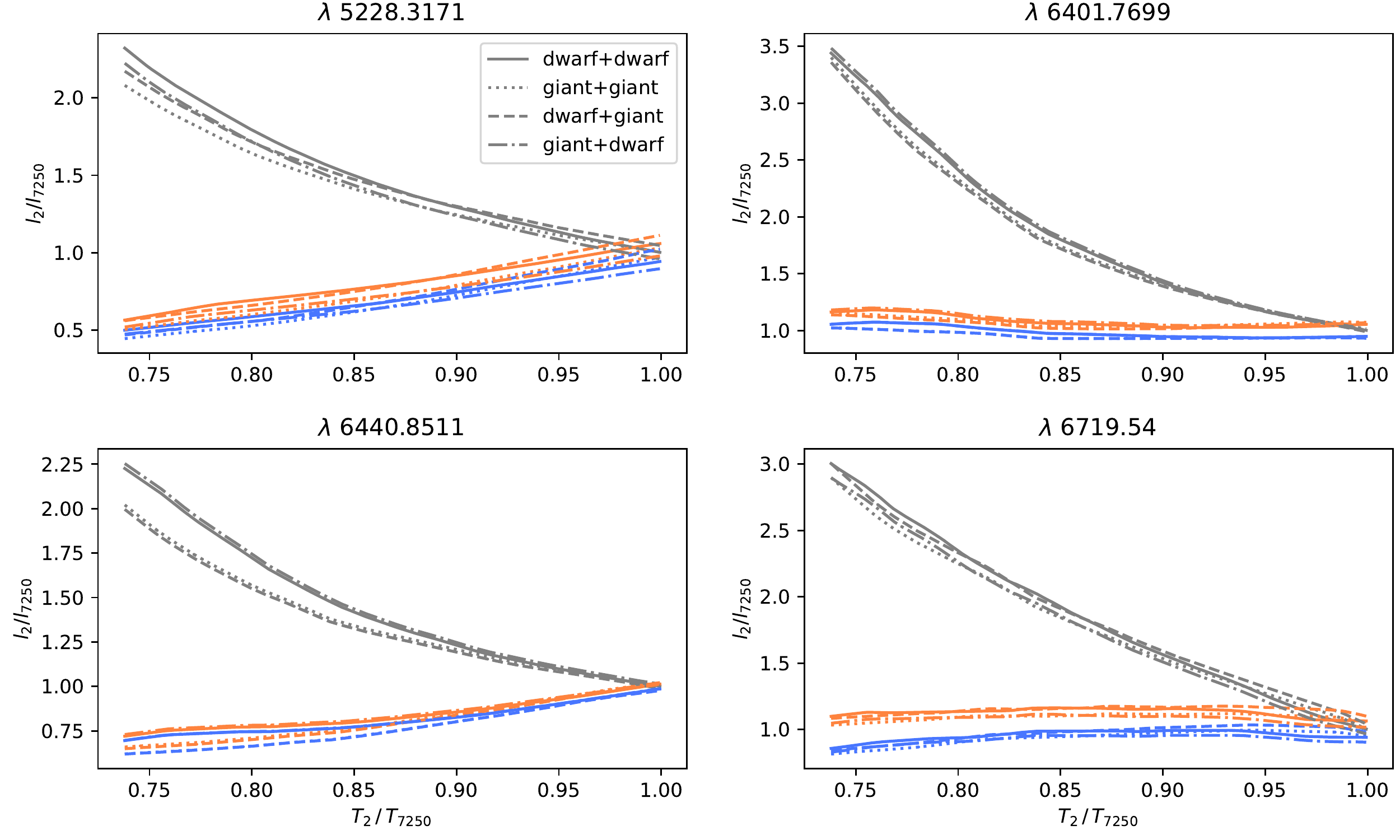}
	\caption{The line intensity ratio between a lower \teff \space star and a 7250 K star. The grey lines represent the ratio between two spectra of single stars, and the colourful lines are the ratio between two star components in the binary spectra. The blue lines mean that the 7250 K components are blue-shifted in SB2 spectra, and the orange lines mean that the 7250 K components are red-shifted. Different line styles stand for different parameter combinations.}
	\label{lineratio-teff}
\end{figure}

\begin{figure}
	\center
	\includegraphics[width=\columnwidth, angle=0]{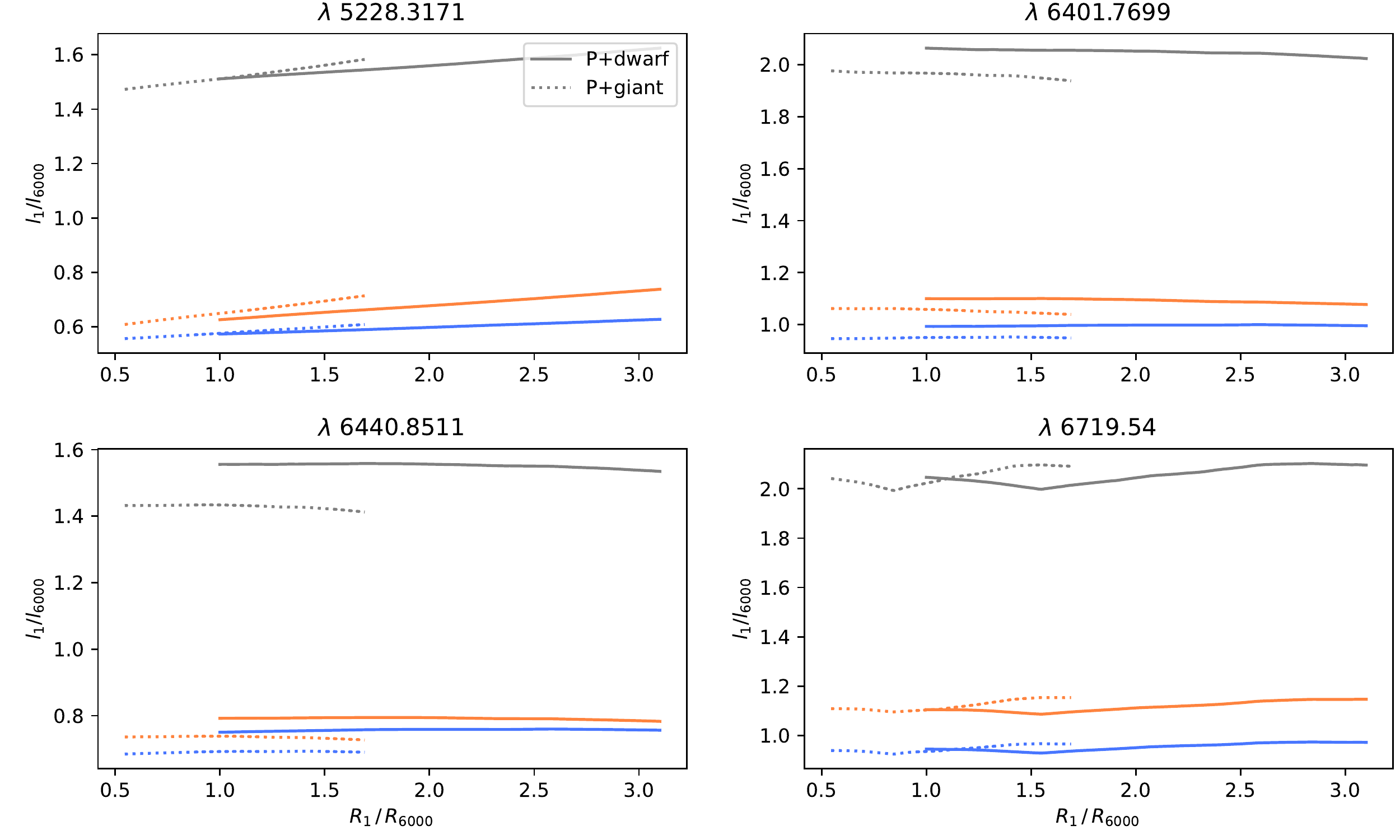}
	\caption{The line intensity ratios change with different radii. The grey lines represent the ratio between two spectra of single stars, and the colourful lines are the ratio between two star components in the binary spectra. The blue lines mean that the 7250 K components are blue-shifted in SB2 spectra, and the orange lines mean that the 7250 K components are red-shifted. Different line styles stand for different parameter combinations.}
	\label{lineratio-R}
\end{figure}

\begin{figure*}
	\center
	\includegraphics[width=16.0 cm, angle=0]{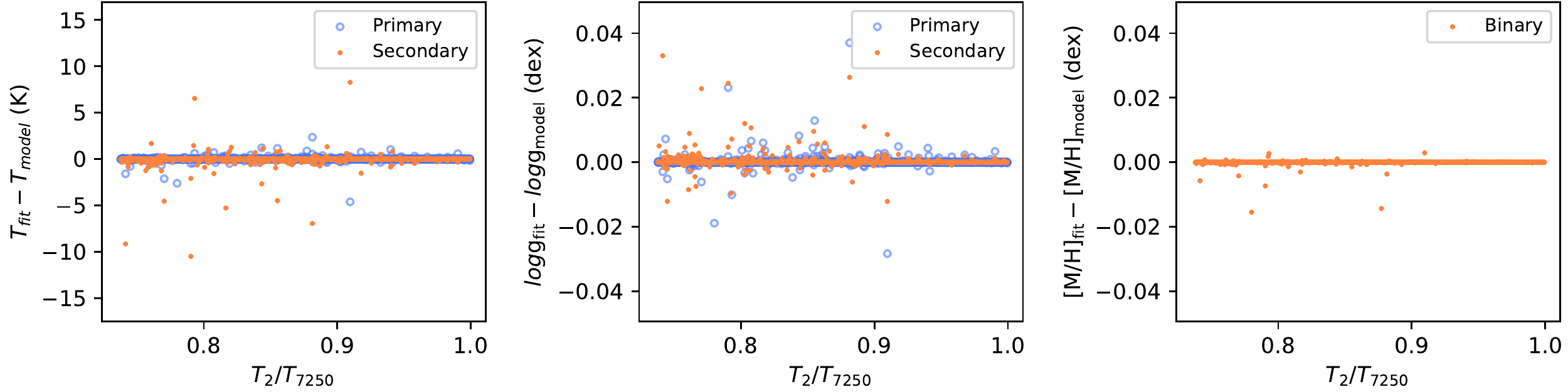}
	\caption{Parameter differences between the measured values and the theoretical values. Blue circles in $\Delta T_{eff}$ and $\Delta \log \, g$ panels show the differences between the primary stars, and orange dots show the differences between the secondary stars. One metallicity is fitted for each binary system, so the $\Delta \rm [M/H]$ panels contain only orange points. All the binaries in this figure have the primary component blue-shifted in the SB2 spectra.}
	\label{dparamdwarf}
\end{figure*}

\begin{figure*}
	\center
	\includegraphics[width=16.0 cm, angle=0]{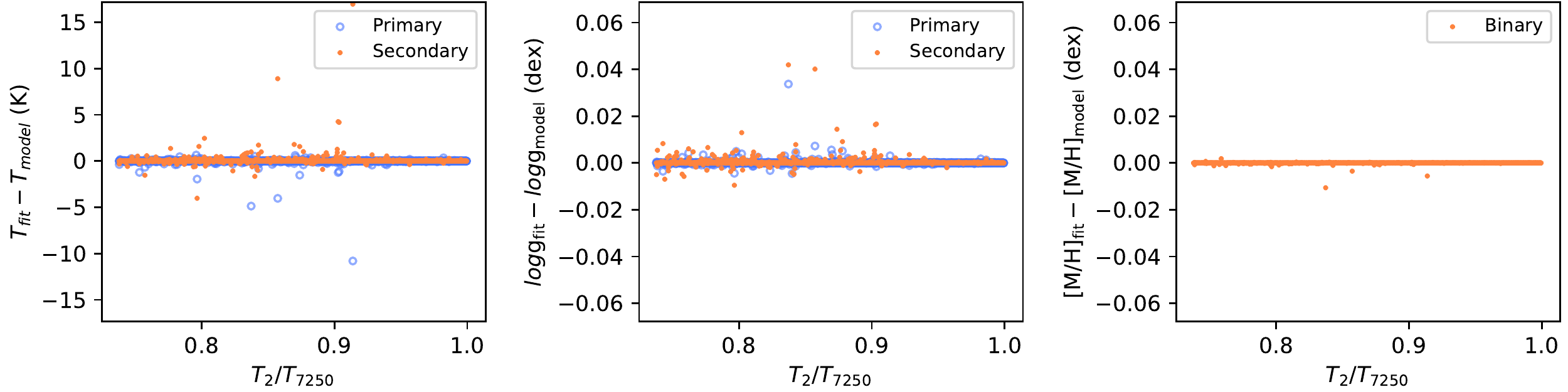}
	\caption{Parameter differences between the measured values and the theoretical values. Colours and symbols have the same meaning as in Figure~\ref{dparamdwarf}. All the binaries in this figure have the primary component red-shifted in the SB2 spectra.}
	\label{dparamgiant}
\end{figure*}

In this subsection, we analyse the impact of the continuum in detail because the pseudo-normalised spectra are used in the fitting, and the line intensities of the normalised spectra will be significantly affected by the continuum. In an SB2 spectrum, the more luminous star contributes more to the continuum and will cause more impact on line intensity after normalisation. The superposition of two continua causes the relative line intensity ratio between two stars in a binary spectrum to be lower than the ratio between two single star spectra. We synthesised a series of SB2 spectra and fitted the double-peaked lines with the double-Gaussian function to see how the effect caused by luminosity and continuum varies with \teff \space and stellar radius. Figure~\ref{teffgra} shows the lines we chose to obtain the line intensity ratios. The upper panel of Figure~\ref{teffgra} shows the gradient of line intensity differences of four \teff \space gaps: $7000 - 50/100/200/500$ K. The differences were calculated after each theoretical spectrum was normalised using the pseudo-continuum. We derived the gradient to find the narrow single lines that are more sensitive to effective temperature changes. If a spectral line is sensitive to \teff change, the gradient shown in the upper panel of Figure~\ref{teffgra} will be higher. The lower panel contains the related theoretical spectra in the LAMOST MRS wavelength and resolving power. Four narrow single spectral lines that are sensitive to \teff \space changing and are less contaminated by nearby lines were selected, and they are marked by the grey dashed lines. Then we generated a series of SB2 spectra using the same model spectra as in our fitting procedures. The chosen stellar parameter ranges that are commonly contained in the LAMOST MRS DR7 data (Figure~\ref{lasp-param}):

a. The metallicity is fixed to be -0.5 dex to reduce the fitting time.

b. To analyse the effect from different effective temperatures, we at first set the primary star with $T_{\rm eff}=7250 $ K (\teff \space for an F0 dwarf star) and let the secondary star have a \teff \space various from 5350 K (\teff \space for a G9 dwarf star) to 7250 K, with a step of 5 K. Then we choose \logg \space equals 4.25 dex and 3.80 dex to represent the dwarf star and the sub-giant star, respectively. According to the MESA Isochrones \& Stellar Tracks (MIST, \cite{MIST0}, \cite{MIST1}), when the primary star is a 7250 K dwarf, it has a radius of 1.25 R$_{\bigodot}$, and if the star is a 7250 K sub-giant, the radii is set to be 2.2 R$_{\bigodot}$. As for the secondary star, we also set the radius to be 1.2 R$_{\bigodot}$ and 2.2 R$_{\bigodot}$ for dwarf stars and sub-giant stars, respectively. We fix the radius of the secondary star to reduce the affection of the stellar radius on the luminosity. The set-up is not always physical but is reasonable in most cases. Next, we generate the double-lined features with $\Delta RV$ between the two component stars equals 100 \kms \space, which is twice the $\Delta RV$ detection limitation of the LAMOST MRS SB2 spectra, to get better spectral line fitting. The SB2 spectra contain four combinations: dwarf + dwarf, sub-giant + sub-giant, dwarf + sub-giant, and sub-giant + dwarf for the primary and the secondary star, respectively. We have 1520 parameter combinations to investigate the \teff \space various affection.

c. We also generated SB2 spectra with various radii to analyse the effect of the stellar radius. The parameter ranges are also from the MIST, although some specific combinations are not physical. To have a more prominent affection, we set the radius of the 7250 K primary star to various from 1.2 R$_{\bigodot}$ to 3.7 R$_{\bigodot}$ with a step of 0.02 R$_{\bigodot}$, and the related surface gravity is between 3.5 dex and 4.3 dex. The secondary star has a \teff \space of 6200 K (for a G0 dwarf star). The \logg \space of a dwarf secondary star is 4.25 dex (R$=1.2$ R$_{\bigodot}$) and of a giant secondary is 3.8 dex (R$=2.2$ R$_{\bigodot}$), the same as in the \teff \space various occasion. The combinations to generate binary spectra are 7250 K primary star + dwarf secondary star and 7250 K primary star + giant secondary star. We have 254 parameter combinations for the radius-changing affection.

d. In different orbital phases, the relative motions between the two component stars make the continuum of the more luminous star blue or red shift in the binary spectra. To analyse if the blue or red-shifted continuum has different effects on the line intensities, the synthesised SB2 spectra have two kinds of radial velocity settings: the primary star has a radial velocity of -50 \kms \space and +50 \kms. $\Delta RV$ between the two components is 100 \kms. This procedure doubles the SB2 spectral number, so we finally have 3548 SB2 spectra for the line intensity ratio analysis, including 3040 spectra with various \teff \space and 508 with various radii. Figure~\ref{lineratio-teff} shows how the line ratios change with different \teff, and Figure~\ref{lineratio-R} shows the changes with various radii.

e. With the same atmospheric parameters and stellar radius, but $\Delta RV$ varies from 40 \kms \space to 120 \kms \space randomly, we generated another 3548 SB2 spectra to evaluate the parameter fitting accuracy of our method. Figure~\ref{dparamdwarf} and Figure~\ref{dparamgiant} show the differences between the theoretical parameters and the fitting parameters.

In Figure~\ref{lineratio-teff}, the line ratios of different spectral lines various with \teff \space are shown in separated panels. In each panel, different line styles represent different parameter combinations. Grey lines are the line intensity ratios between two spectra of single stars. The blue lines represent that the $7250 $ K components are blue-shifted in the SB2 spectra, and the orange lines mean the components are at the red-shifted phases. In the normalised single star spectra, the relative line intensities of lower \teff \space stars are stronger than that of the stars have \teff$=7250 $ K. But in an SB2 spectrum, the superposition of two continua will reduce the relative line intensities after normalisation. The blue and the orange lines mean that the line intensities of the less luminous star are reduced more than that of the other.

Figure~\ref{lineratio-R} shows how the line intensity ratios change with different radii. Line styles and colours in this figure have the same meaning as in Figure~\ref{lineratio-teff}. The line intensity ratios don't vary a lot with stellar radius may be caused by our narrow radius selection range. The differences between the grey and the blue/orange lines also mean that the superposed continuum has more affection for the line intensities of the less luminous star.

Although the line intensities of the less luminous star are significantly reduced in the binary spectra, the blue and orange lines that have the same line style in Figure~\ref{lineratio-teff} and Figure~\ref{lineratio-R} mean that the line intensity decrease does not vary a lot in different orbital phases. The atmospheric parameters derived in different phases (and with various $\Delta RV$) should be consistent as long as the components can be detected. Figure~\ref{dparamdwarf} shows the atmospheric parameter differences of the synthetic SB2 spectra with the primary stars are in blue-shifted phases, and Figure~\ref{dparamgiant} is the parameter differences when the primary stars are at red-shifted phases. Each panel contains the parameter from the fitting results of both the primary and the secondary star. The figures show that if the stellar radius and luminosity contribution are known, the fitting results will have a very low-level uncertainties for spectra without noises. So we do not consider the method uncertainties in the application on the observed SB2 spectra.

%__________________________________________________________________

\section{Examples: Atmospheric Parameters of Two SB2 Eclipsing Systems}
\label{sect:parameters}

In this section, we give two observation examples of eclipsing binaries to show the ability of our method. The examples are chosen randomly because the method is designed for binary stars that match our criteria in Sect~\ref{sect:method}. Both of the binaries have at least three SB2 spectra in LAMOST MRS DR7 and have light curves from Kepler/K2 or TESS.

\subsection{TIC 63209649}
\label{subsec:eg02}		

\begin{figure}
	\center
	\includegraphics[width=\columnwidth, angle=0]{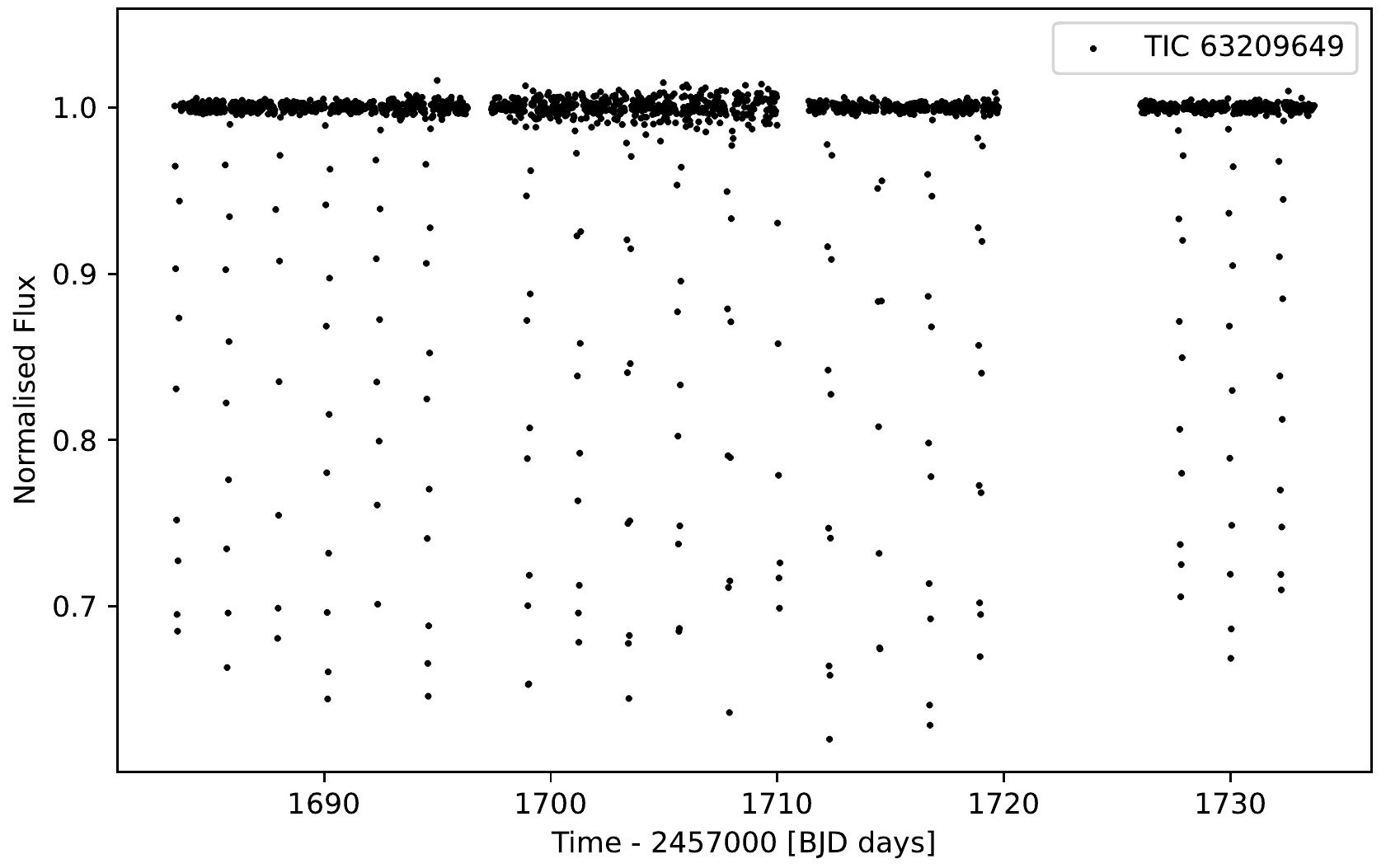}
	\caption{Normalised TESS light curve of TIC 63209649.}
	\label{eg02lc}
\end{figure}

\begin{figure}
	\center
	\includegraphics[width=\columnwidth, angle=0]{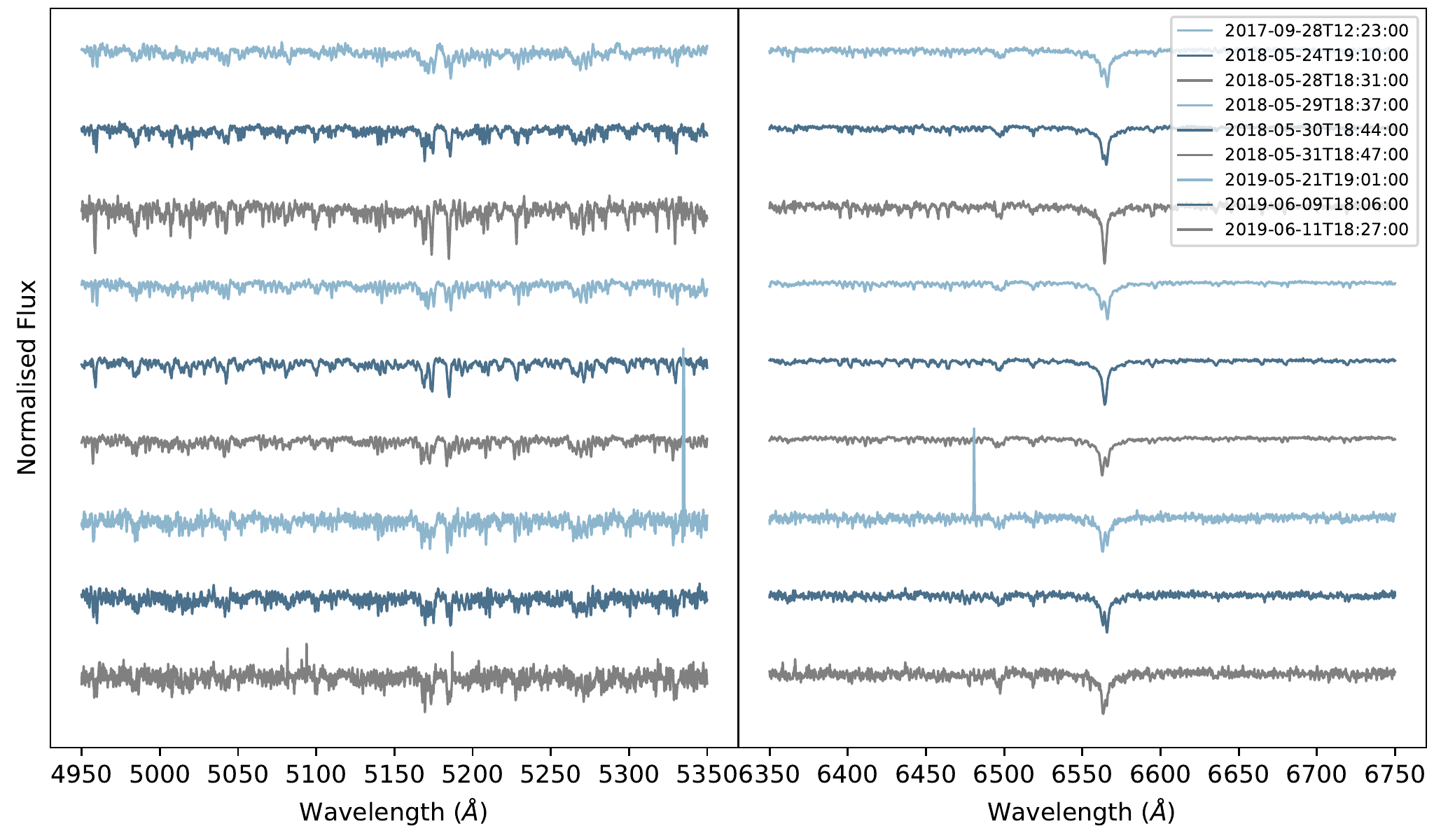}
	\caption{Pseudo-continuum normalised LAMOST MRS multi-epoch spectra of TIC 63209649. The spectra are shown in observed time order.}
	\label{eg02spec}
\end{figure}

\begin{figure}
	\center
	\includegraphics[width=\columnwidth, angle=0]{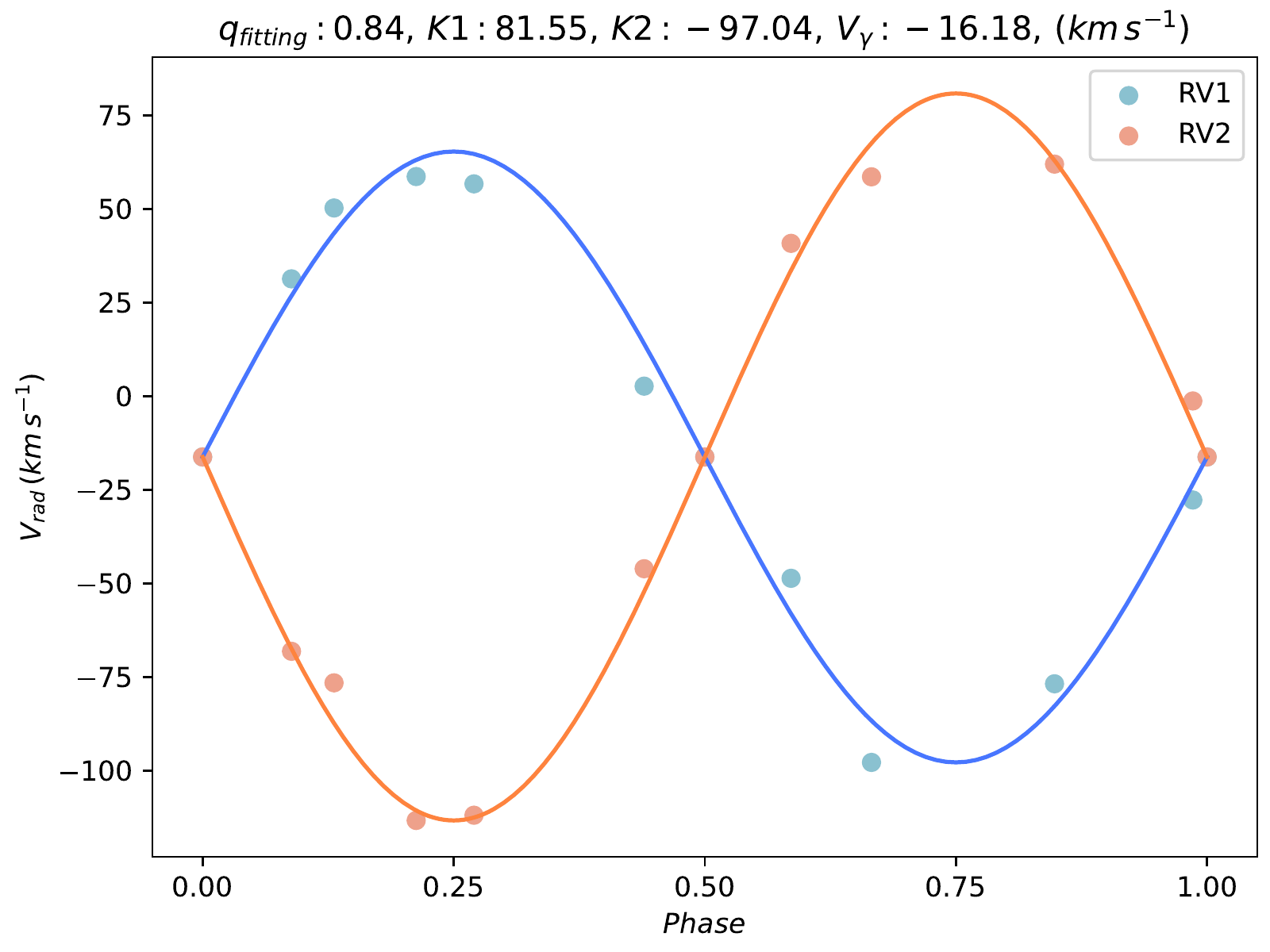}
	\caption{The reconstructed radial velocity curves of TIC 63209649. Dots are the RVs measured by the spectra except for three dots in the phases 0, 0.5 and 1 that are set manually to be $ \gamma $. Lines are the reconstructed curves. Blue represents the higher mass star and orange represents the lower mass star, respectively.}
	\label{eg02rv}
\end{figure}

\begin{figure}
	\center
	\includegraphics[width=5.5cm, angle=270]{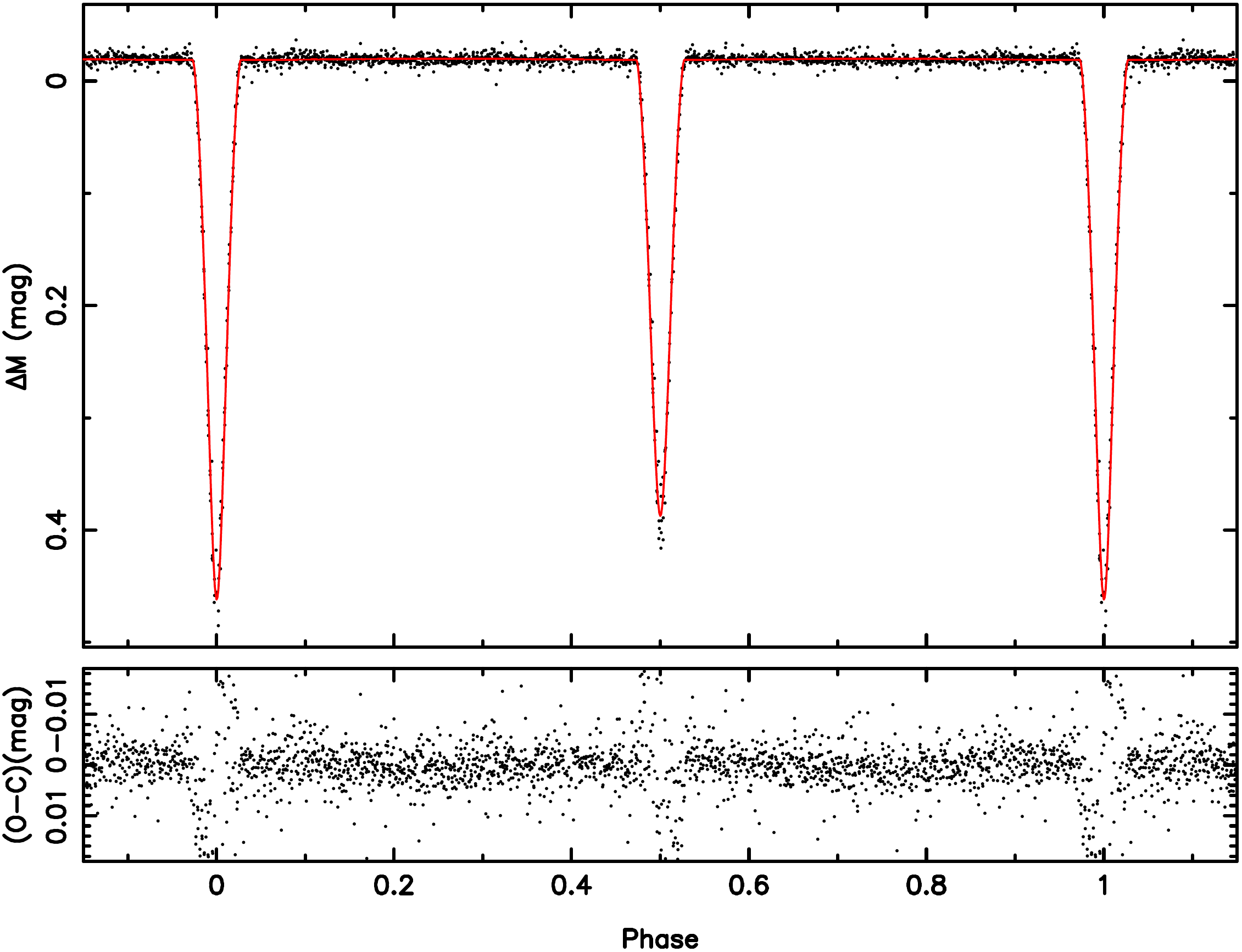}
	\caption{WD fitting result of the TIC 63209649 light curve. Black dots in the upper panel are the observed data and the red line is the best fitting model. The lower panel shows the fitting residuals.}
	\label{eg02orbit}
\end{figure}

\begin{figure*}
	\center
	\includegraphics[width=16.0 cm, angle=0]{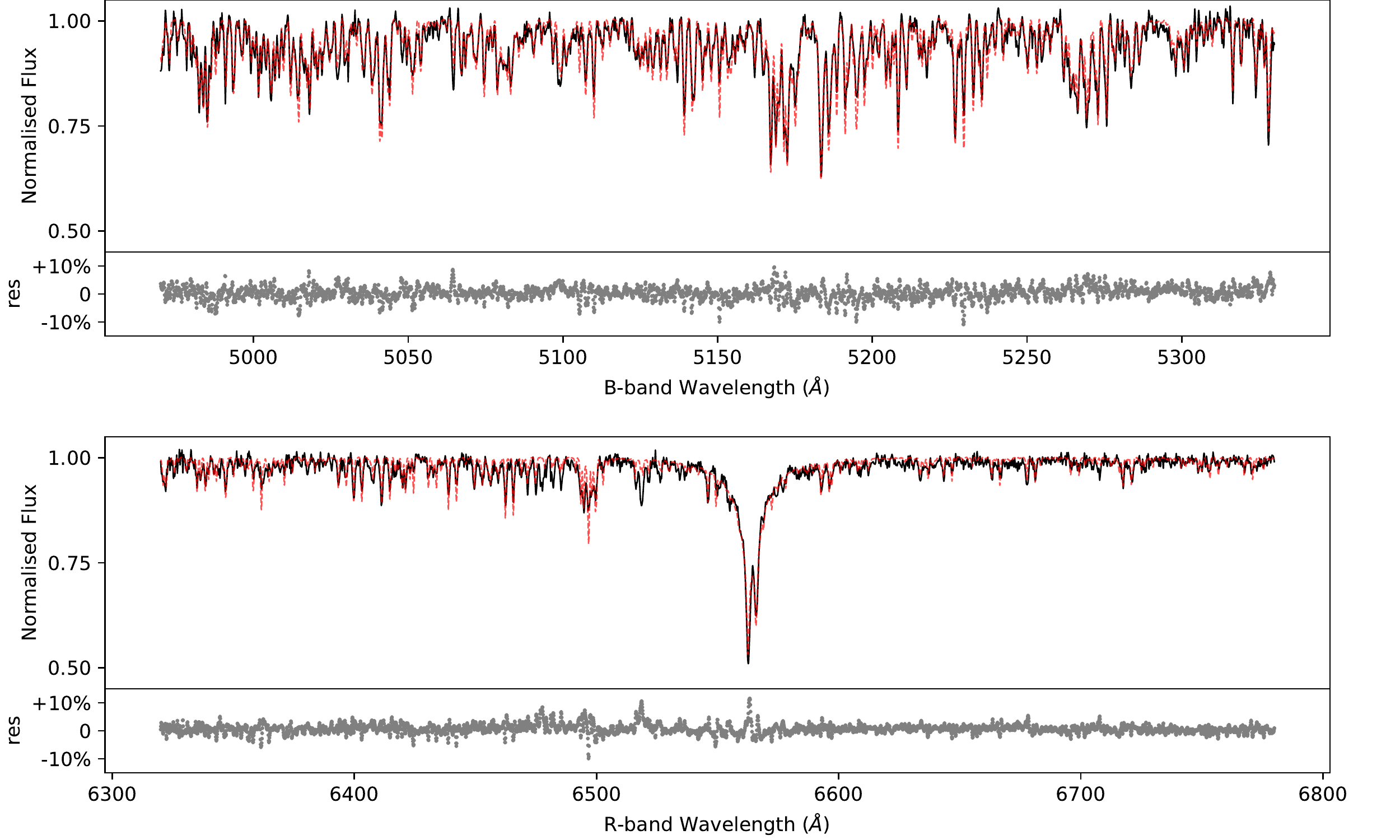}
	\caption{A best-fitting example of the TIC 63209649 SB2 spectrum. Black lines are the LAMOST MRS spectra and red lines are the best fitting synthetic SB2 model. Grey dots represent the fitting residuals.}
	\label{eg02fitting}
\end{figure*}

TIC 63209649 (or say KIC 8301013. TESS light curves are used in this work, so we choose the TIC number) is a detached eclipsing binary system with nine successful epochs of spectra in the LAMOST MRS DR7. Figure~\ref{eg02lc}, \ref{eg02spec} and \ref{eg02rv} show the observation data. Figure~\ref{eg02lc} is the normalised TESS light curve. This system has a period of about 4.43 days. Figure~\ref{eg02spec} shows the normalised LAMOST MRS spectra of nine epochs. The spectra are listed in the observed time sequence, and the observation time in UTC format is shown in the upper right legend. Figure~\ref{eg02rv} shows the radial velocities of component stars in phase view.  In order to reconstruct the RV curve, we firstly use \emph{the Payne} and our binary spectral model to fit each binary spectrum. The centres of strong lines can be fitted correctly to obtain the radial velocities of each star. Then we use $q=\frac{rv_1-\gamma}{\gamma-rv_2}$ to fit $q$ and $\gamma$. $rv_1$ and $rv_2$ are the radial velocities of the primary star and the secondary star folded in the phase order, respectively. Except for the three dots in phases 0, 0.5 and 1, other dots in Figure~\ref{eg02rv} are the measured RVs. The three dots are set manually to be $ \gamma $. Finally, we reconstruct the RV curve using cosine functions. The standard deviation between the measured RVs and the best fitting lines is $5.08 $ \kms \space for RV1 and $ 6.66 $ \kms \space for RV2. Note that the fitting may be rough at the near-eclipsing phases. Combining the light curve and the radial velocity curves, we derived all the orbital parameters using the WD modelling. Figure~\ref{eg02orbit} is the WD fitting result. Black dots in the upper panel represent the folded light curve, and the red line is the best fitting model. Dots in the lower panel are the residuals. The orbital parameters from the light curve solution are listed in Table~\ref{tab:eg02lcparam}. Combining the orbital parameters, we generated the synthetic SB2 model spectra according to equation~\ref{equ-area}, and fitted seven multi-epoch SB2 spectra of TIC 63209649. Figure~\ref{eg02fitting} shows the best fitting result of one of the SB2 epochs. Black lines are the observed spectra of blue (upper panel) and red (lower panel) bands, and the red lines represent the best fitting spectra. Grey dots indicate the percentage of the residual in the observed data. Most of the residuals are below ten per cent of the observed spectra, except for residuals of some strong lines like the Mg I triplet, $ \lambda $5167, 72, 83 and the H$ \alpha $ line. In wavebands with dense spectral lines, for example, $ \lambda $5175$ \sim $5180, and $ \lambda $5280$ \sim $5285, line mixing causes the fitting residuals to become a bit higher than the relatively line-less waveband. Two of the nine epochs were observed near eclipsing phases and did not show clear double-lined features. We present the spectral fitting results in Appendix~\ref{appendix-b} and give a brief discussion.

\begin{figure}
	\center
	\includegraphics[width=\columnwidth, angle=0]{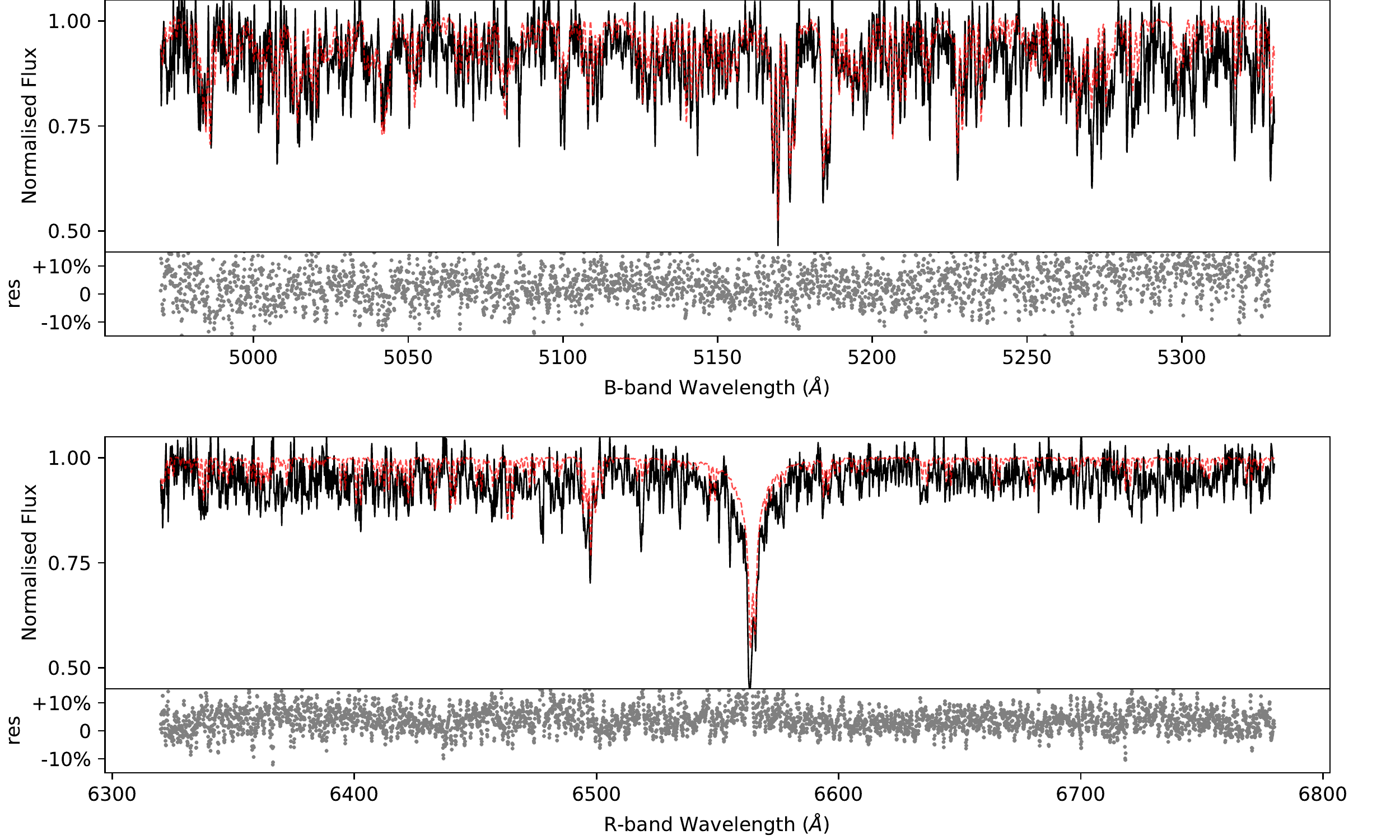}
	\caption{A fitting example of a TIC 63209649 spectrum with a low signal-to-noise ratio. Colours and symbols have the same meaning as in Figure~\ref{eg02fitting}.}
	\label{eg02sn}
\end{figure}

Derived atmospheric parameters of all the SB2 spectra of TIC 63209649 are listed in Table~\ref{tab-eg02specparam} in order of observation time. The orbital phases are shown in the second column. $T_{\rm eff1}$ and log\,$\it g_{\rm 1}$ are the parameters of the hotter star, and the parameters of the cooler one are $T_{\rm eff2}$ and log\,$\it g_{\rm 2}$. The metallicity of the binary system is \feh. The parameters of the first four epochs have a good consistency. But the results of the last three epochs are different, and they are labelled with asterisks in the first column. We checked the spectra of these three epochs and found that a poor signal-to-noise ratio might be the reason for this inconsistency. Figure~\ref{eg02sn} shows the fitting result of the last spectrum (MJD=58645) in Table~\ref{tab-eg02specparam}. The noise in this spectrum is higher compared with that in Figure~\ref{eg02fitting}. We thus excluded this spectrum as well as the two previous noisy spectra before parameter averaging. Atmospheric parameters in the last row of Table~\ref{tab-eg02specparam} are the mean values of the adopted epoch results. We take the standard deviations as the measurement errors and list them in the parentheses. The errors are at the same level as the errors of LASP parameters of the single stars (\cite{MRSvalue2021Chen}).

\begin{table}
	\centering
	\caption{Orbital parameters of TIC 63209649.}
	
	\begin{tabular}{p{0.25\linewidth}lll}
		\hline
		\noalign{\smallskip}
		Parameter  & Primary & System & Secondary  \\
		\noalign{\smallskip}
		\hline
		\noalign{\smallskip}
		$t_{conj} \ (d)$         &       & 2458685.7373    &      \\
		
		$P \ (d)$                 &       & 4.4275    &      \\
		
		$\gamma \ (km \, s^{-1})$ &       & -16.1754    &      \\
		
		$q$                       &       & 0.8403    &      \\
		
		$a\sin i \ (R_{\bigodot}) $ &       & 16.11$\pm$0.053  &      \\
		
		$i \ (^{\circ})$          &       & 87.44$\pm$0.03  &      \\
		
		$e$                       &       & 0.0299 $\pm$0.0177 &     \\
		
		$\omega \ (^{\circ})$     &       & 90.25 $\pm$0.22 &      \\
		
		$r_{\rm T}$               &       & 0.9700 &      \\
		
		$R \ (R_{\bigodot})$      & 1.5093&        & 1.2297\\
		
		$M \ (M_{\bigodot})$      & 1.4305&        & 1.2016\\
		
		log\,$\it g \ $ (dex)       & 4.24  &        & 4.34\\
		\noalign{\smallskip}
		\hline
	\end{tabular}
	\label{tab:eg02lcparam}
\end{table}

\begin{table*}
	\centering
	\caption{Atmospheric parameters of the TIC 63209649 component stars. Lines with asterisks are excluded before deriving the final parameters because of the low signal-to-noise ratios.}
	
	\begin{tabular}{p{0.25\linewidth}llllll}
		\hline
		\noalign{\smallskip}
		MJD  & Phase & $T_{\rm eff1}$ (K) & log\,$\it g_1$ (dex) & $T_{\rm eff2}$ (K) & log\,$\it g_2$ (dex) & $\rm [M/H]$ (dex)  \\
		\noalign{\smallskip}
		\hline
		\noalign{\smallskip}
		58024 & 0.2702   & 6724.98    & 4.18        & 5946.48    & 3.78        & -0.34     \\
		58262 & 0.0885   & 6596.99    & 4.08        & 5962.63    & 3.83        & -0.07     \\
		58267 & 0.2126   & 6614.74    & 3.94        & 5880.60    & 3.73        & -0.37     \\
		58269 & 0.6659   & 6636.17    & 4.30        & 5879.86    & 3.79        & -0.14     \\
		58624 $\ast$ & 0.8481   & 6386.38    & 3.56        & 5966.20    & 4.45        & 0.10     \\
		58643 $\ast$ & 0.1308   & 6232.98    & 3.69        & 5678.84    & 4.25        & -0.28     \\
		58645 $\ast$ & 0.5858   & 6140.11    & 3.58        & 5817.86    & 3.56        & -0.25     \\
		Mean(Std) & -   & 6643.22(49.20)    & 4.13(0.13)        & 5917.39(37.60)    & 3.78(0.04)        & -0.23(0.13)     \\
		\noalign{\smallskip}
		\hline
	\end{tabular}
	\label{tab-eg02specparam}
\end{table*}

\subsection{EPIC 247529791}
\label{subsec:eg01}

\begin{figure}
	\center
	\includegraphics[width=\columnwidth, angle=0]{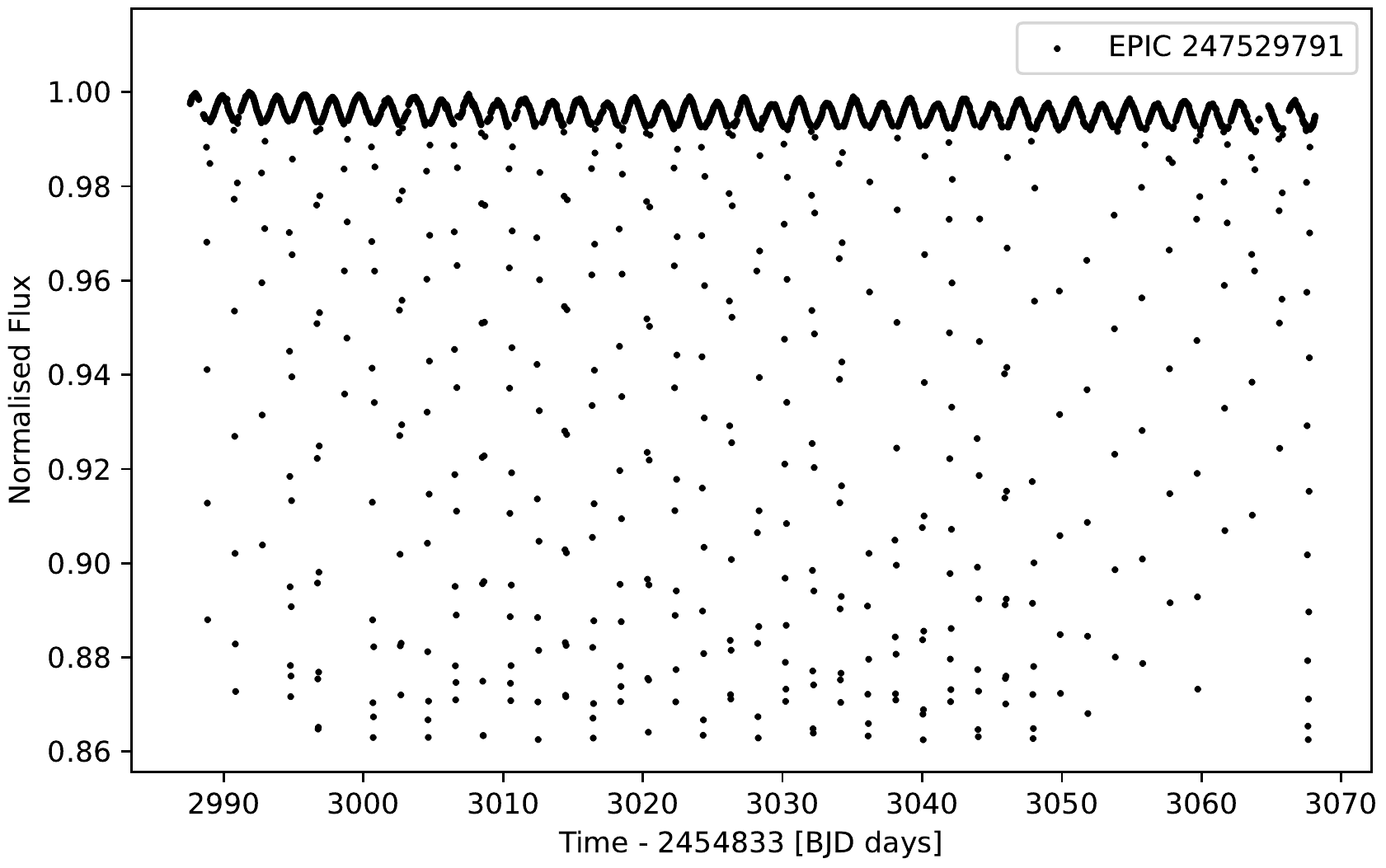}
	\caption{Normalised K2 light curve of EPIC 247529791.}
	\label{eg01lc}
\end{figure}

\begin{figure}
	\center
	\includegraphics[width=\columnwidth, angle=0]{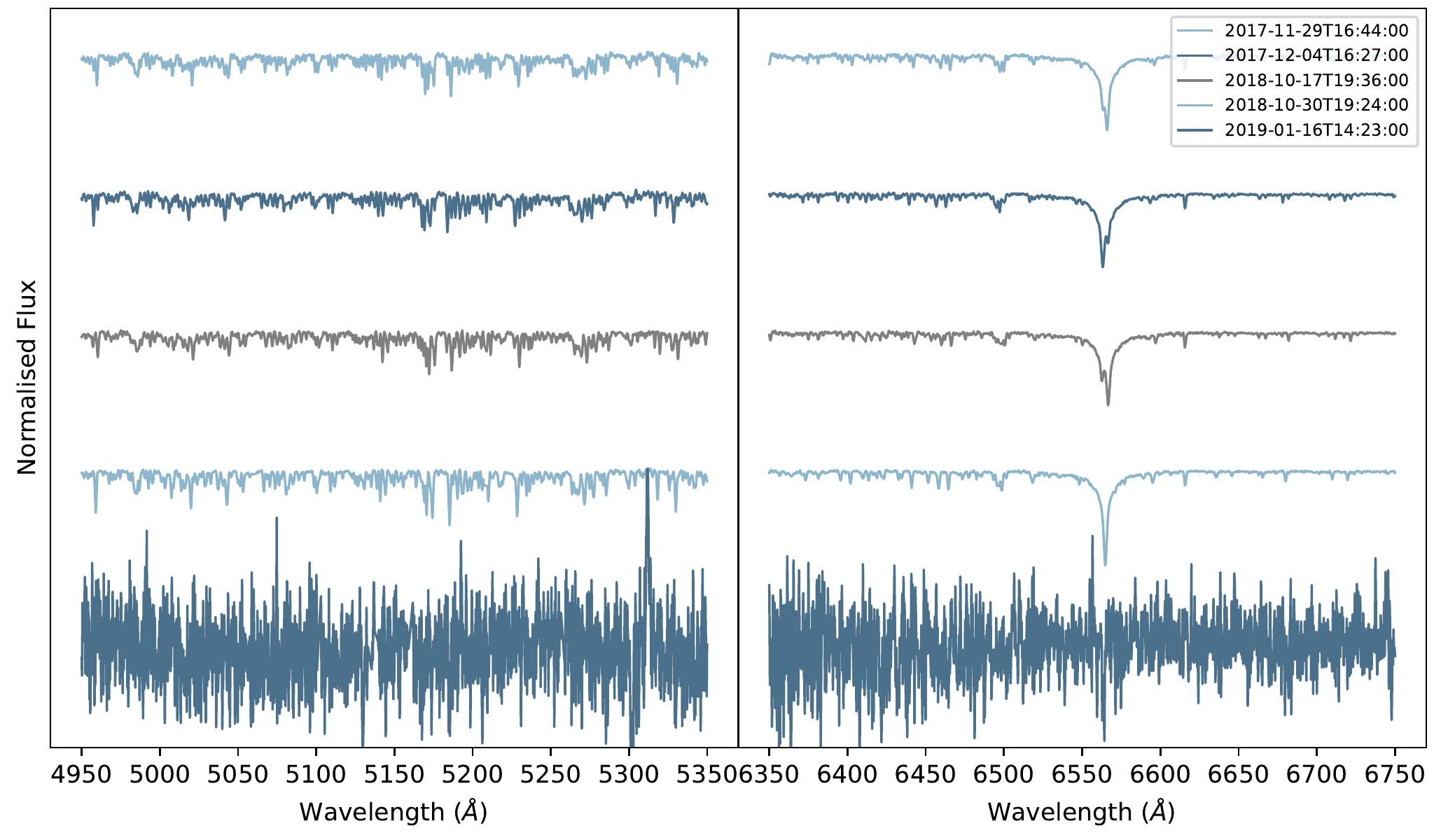}
	\caption{Pseudo-continuum normalised LAMOST MRS multi-epoch spectra of EPIC 247529791. The spectra are shown in observed time order.}
	\label{eg01spec}
\end{figure}

\begin{figure}
	\center
	\includegraphics[width=\columnwidth, angle=0]{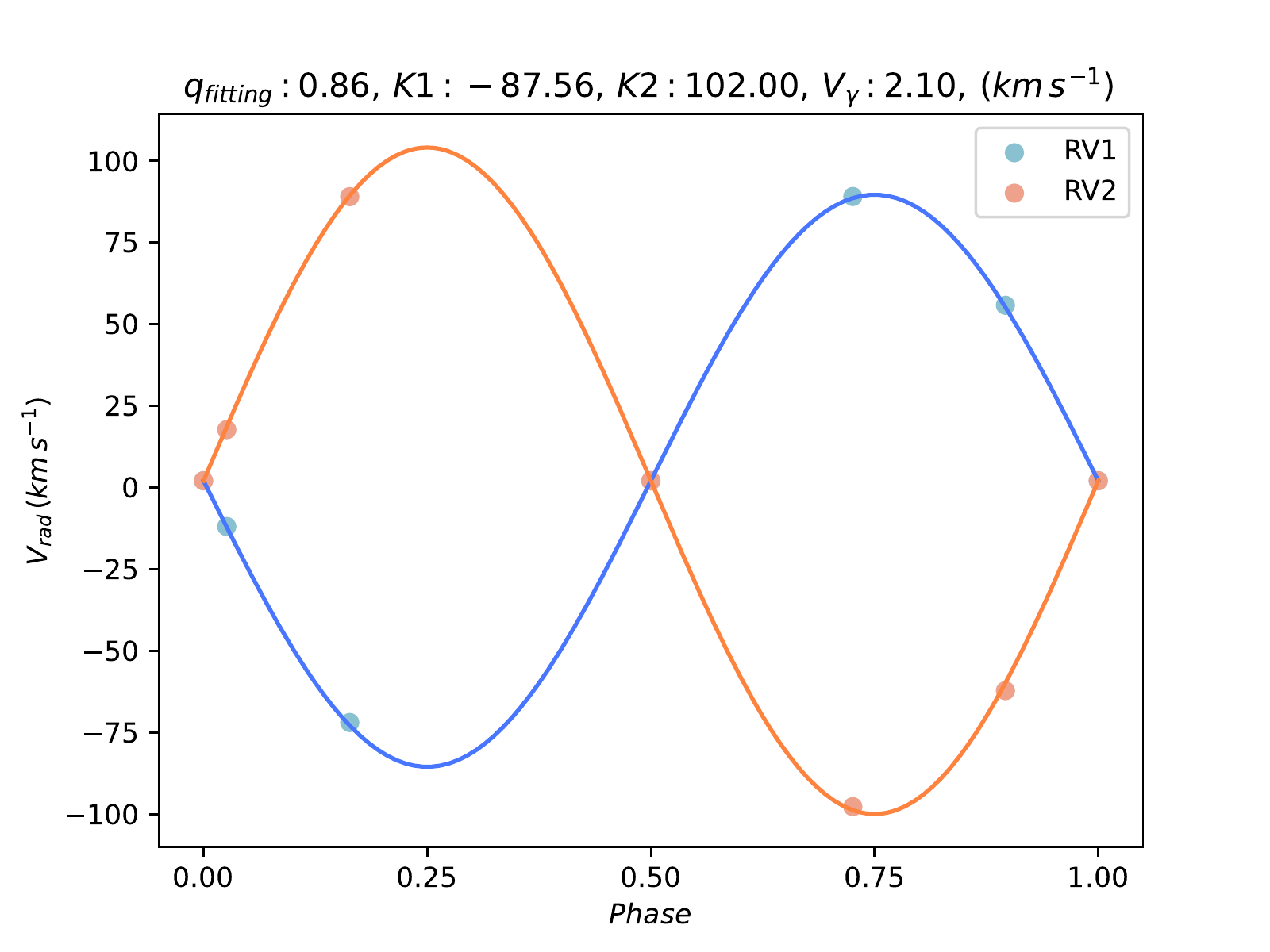}
	\caption{The reconstructed radial velocity curves of EPIC 247529791. Dots are the RVs measured by the spectra except for three dots in the phases 0, 0.5 and 1 that are set manually to be $ \gamma $. Lines are the reconstructed curves. Blue represent the higher mass star and orange represent the lower mass star, respectively.}
	\label{eg01rv}
\end{figure}

\begin{figure}
	\center
	\includegraphics[width=5.5 cm, angle=270]{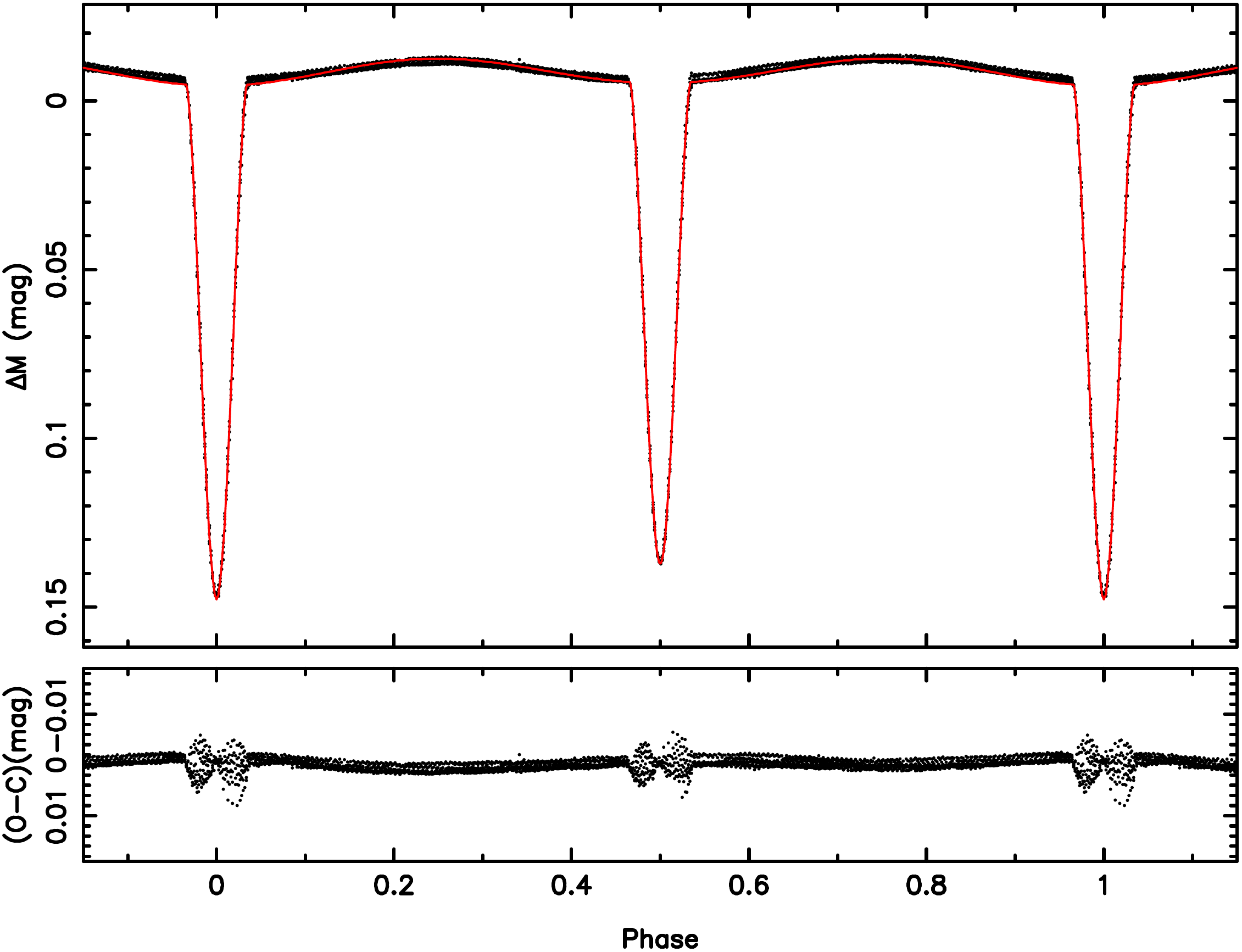}
	\caption{WD fitting result of the EPIC 247529791 light curve. Black dots in the upper panel are the observed data and the red line is the best fitting model. The lower panel shows the fitting residuals.}
	\label{eg01orbit}
\end{figure}

\begin{figure*}
	\center
	\includegraphics[width=16.0cm, angle=0]{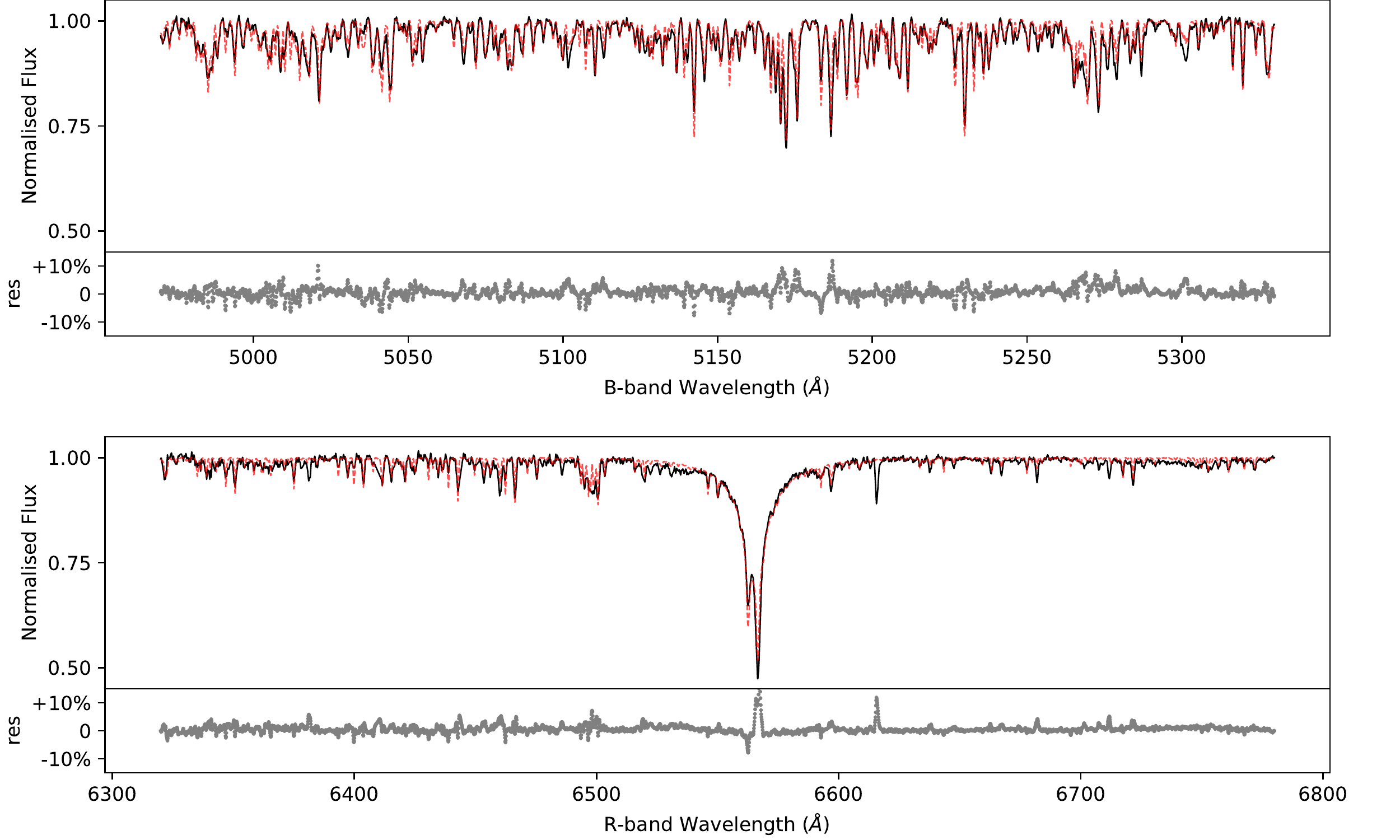}
	\caption{A best-fitting example of the EPIC 247529791 SB2 spectrum. Black lines are the LAMOST MRS spectra and red lines are the best fitting synthetic SB2 model. Grey dots represent the fitting residuals.}
	\label{eg01f1}
\end{figure*}

The K2 target EPIC 247529791 is also a detached eclipsing binary with a nearly circular orbit. Figure~\ref{eg01lc} is the normalised K2 light curve. The orbital period is 3.94 days. Figure~\ref{eg01spec} shows LAMOST MRS spectra of five epochs in time sequence. The latest spectrum has a bad signal-to-noise ratio, so we excluded it in the fitting procedures. Figure~\ref{eg01rv} shows the reconstructed radial velocity curves. The curves are fitted by the same procedures as in Sect~\ref{subsec:eg02} and the symbols and lines have the same meaning as in Figure~\ref{eg02rv}. The standard deviation between the measured RVs and the best fitting lines are $ 0.33 $ \kms \space for RV1 and $ 0.98 $ \kms \space for RV2. The orbital parameters are also derived using WD and Figure~\ref{eg01orbit} is the best fitting model. The orbital parameters are listed in Table~\ref{tab-eg01lcparam}. The model synthesising and spectra fitting is the same as above. Three of the four higher S/N spectra show double-lined features and are used to derive stellar parameters. Figure~\ref{eg01f1} is an example of best-fitting results. In most of the wavelengths, the residuals have the same level as in Figure~\ref{eg02fitting}. The fitting result of the unresolved spectrum is also discussed in Appendix~\ref{appendix-b}.

\begin{table}
	\centering
	\caption{Orbital parameters of EPIC 247529791.}
	
	\begin{tabular}{p{0.25\linewidth}lll}
		\hline
		\noalign{\smallskip}
		Parameter  & Primary & System & Secondary  \\
		\noalign{\smallskip}
		\hline
		\noalign{\smallskip}
		$t_{conj} \ (d)$         &       & 2457821.8943    &      \\
		
		$P \ (d)$                 &       & 3.9365    &      \\
		
		$\gamma \ (km \, s^{-1})$ &       & 2.1000    &      \\
		
		$q$                       &       & 0.8585    &      \\
		
		$a\sin i \ (R_{\bigodot})$  &       & 14.60$\pm$0.023 & \\
		
		$i \ (^{\circ})$          &       & 80.803$\pm$0.006    &      \\
		
		$e$                       &       & 0.0072 $\pm$0.0015 &   \\
		
		$\omega \ (^{\circ})$     &       & 84.6 $\pm$2.9 &  \\
		
		$r_{\rm T}$               &       & 0.9709 &      \\
		
		$R \ (R_{\bigodot})$      & 2.3715&        & 1.5946     \\
		
		$M \ (M_{\bigodot})$      & 1.5576   &            & 1.3395     \\
		
		log\,$\it g \ $ (dex)           & 3.88   &            & 4.16     \\
		\noalign{\smallskip}
		\hline
	\end{tabular}
	\label{tab-eg01lcparam}
\end{table}

Table~\ref{tab-eg01specparam} shows the atmospheric parameters of SB2 spectra fitting. \teff \space and \feh \space derived from different phases have good consistency, but \logg \space of the secondary star can not be derived robustly. Two reasons may explain this phenomenon. First, the lower S/N influences the fitting result a lot. Second, the most prominent line features in the LAMOST MRS spectra are the Mg I triplet, $\lambda$5167, 72, 83 lines in the blue band, and the H$\alpha$ line in the red band. The Mg I triplet lines are not sensitive enough to luminosity (\cite{Mglogg1984}), and their line profiles are easily influenced when $\Delta RV$ between component stars get larger. Although combining the red band in the fitting procedure can improve the \logg \space derivation (\cite{MRSvalue2021Chen}, Figure 11$\sim$12), the wings of the H$\alpha$ line are also significantly affected by the mixing of two components.

\begin{table*}
	\centering	
	\caption{Atmospheric parameters of the EPIC 247529791 component stars.}
	
	\begin{tabular}{p{0.25\linewidth}llllll}
		\hline
		\noalign{\smallskip}
		MJD  & Phase & $T_{\rm eff1}$ (K) & log\,$\it g_1$ (dex) & $T_{\rm eff2}$ (K) & log\,$\it g_2$ (dex) & $\rm [M/H]$ (dex)  \\
		\noalign{\smallskip}
		\hline
		\noalign{\smallskip}
		58086 & 0.8963   & 6949.89    & 3.58        & 6180.68    & 3.73        & -0.46     \\
		58091 & 0.1634   & 6969.50    & 3.47        & 6166.86    & 3.00        & -0.43     \\
		58408 & 0.7257   & 6989.81    & 3.62        & 6092.99    & 3.44        & -0.48     \\
		Mean (Std) & -   & 6969.73 (16.30)& 3.56 (0.06)& 6146.84 (38.50)    & 3.39 (0.30)        & -0.46 (0.02)     \\
		\noalign{\smallskip}
		\hline
	\end{tabular}
	\label{tab-eg01specparam}
\end{table*}

The atmospheric parameters of component stars in the EPIC 247529791  system were also derived by \cite{galahbinary} (hereafter Traven2020). The GALAH DR2 specID of this binary is 17013100180129. In that work, they used photometry data to reconstruct the SED of the binaries and then applied the Bayesian inference and a Monte Carlo Markov chain sampler (\cite{BayesianBookC3, Hogg2010, FMemcee2013, Sharma2017}) to derive atmospheric parameters. The Traven2020 parameters are listed in Table~\ref{tab-eg01galahparam}. Metallicities of the system derived in our work and by Traven2020 have a good consistency. For the warmer component star, our method gets higher \teff \space than Traven2020, while for the cooler star, our \teff \space is lower than the result of Traven2020. The \logg \space of the warmer star obtained by the two methods are similar. The \logg \space of the cooler star in our results is lower than in the Traven2020 results. Furthermore, both \logg \space of the cooler star derived in the two works is lower than that in the light curve solution in Table~\ref{tab-eg01lcparam}.

\begin{table}
	\centering
	\caption{Atmospheric parameters of the EPIC 247529791 derived by Traven2020.}
		
	\begin{tabular}{p{0.5\linewidth}l}
		\hline
		\noalign{\smallskip}
		Parameter              & Value      \\
		\noalign{\smallskip}
		\hline
		\noalign{\smallskip}
		$R_1 \ (R_{\bigodot})$  & 1.65       \\
		
		$T_{\rm eff1}$ (K)     & 6729.99    \\
		
		log\,$\it g_1$ (dex)   & 3.67       \\
		
		$R_2 \ (R_{\bigodot})$  & 2.77       \\
		
		$T_{\rm eff2}$ (K)     & 6431.46    \\
		
		log\,$\it g_2$ (dex)   & 3.80       \\
		
		[M/H] (dex)        &  -0.42     \\
		\noalign{\smallskip}
		\hline
	\end{tabular}
	\label{tab-eg01galahparam}
\end{table}

Different resolving power and fitting methods may cause the inconsistency of the derived parameters. (1) The LAMOST MRS with R$\sim$7500 has a lower resolution than GALAH (R$\sim$28000). Line mixing caused by the Doppler shift of the component stars makes recognising features of the fainter star more difficult, and makes deriving atmospheric parameters imprecisely in the lower resolution spectra. (2) We use the Kurucz theoretical spectra as our fitting model, while Traven2020 adopted a series of observed spectra to be the fitting template, same as the GALAH official pipeline (\cite{GALAHpipeline2017}). (3) The binary spectra models are produced in different ways. In our work, we synthesise the binary spectra with the light curve-derived radii and the phase-based projection areas. The luminosity contribution depends on the real size and \teff \space ratio of the component stars. Traven2020 generated the synthetic SED of two single stars using the Kurucz ATLAS9 models and produced the SED of the binary system taking the component star radii as the independent variables. The generated binary SED and the radii were restricted by the apparent magnitudes from AAVSO Photometric All-Sky Survey – APASS (\cite{AAVSO2016}), Gaia DR2 (\cite{GaiaDR22018photo}), Two Micron All Sky Survey – 2MASS (\cite{TwoMASS2006}), and Wide-field Infrared Survey Explorer – WISE (\cite{WISE2010}).

Except for the difference between the two methods, we suggest other two factors that may cause the inconsistency. The two factors may also be the disadvantages of deriving atmospheric parameters using the SB2 spectra. 

First, the effect of mixing and continuum contamination on the line depth influence \teff \space measurement. Theoretically, the prime criteria to derive the effective temperature for a main-sequence F-type star are the strength and profiles of the hydrogen lines (\cite{Gray2009}). But the LAMOST MRS waveband contains only the H$\alpha$ line. Due to its depth and broad width, the H$\alpha$ line is easily influenced by the mixing of two spectra. In this case, the metal lines become the main features to derive \teff. But metal lines are shallower than the H$\alpha$ lines; the absolute strength is more easily affected by the continuum. And the relative line intensity ratio is constrained by the luminosity contribution. When fitting the SB2 spectrum, the more luminous star component occupies more weight in fitting the best model. This makes the measured \teff \space of the hotter star higher than the real value. The two reasons say the relative line intensity change and the higher measured \teff \space of the hotter star cause the \teff \space ratio to get smaller in the fitting iteration, which makes the \teff \space measurement of the cooler star to be lower. Absolute flux calibration may help to reduce continuum contamination.

Second, the surface gravity of the component stars by fitting the SB2 spectra are underestimated. Since the most prominent spectral lines in the LAMOST MRS waveband are the MgI triplet $\lambda \lambda$ 5167, 72, 83 and the H$\alpha$, the surface gravity of most of the A- to G-type MS stars can not be derived precisely (\cite{MRSvalue2021Chen}). Wings of the hydrogen lines are luminosity sensitive for early A-type stars, and the MgI triplet is for late G- to mid-K-type stars (\cite{PASP1984}). Furthermore, the micro-turbulent velocity plays an important role in determining surface gravity for the F-, G-  and even K-type stars. For the mid-F-type dwarfs, a micro-turbulent velocity of about two \kms \space affects the luminosity criteria (\cite{microturbulence2001}). The spectral lines in a single star spectrum are broader and stronger due to the micro-turbulence. In an SB2 spectrum, Doppler shift and line mixing cause the double-peaked lines to be broader and stronger than the lines in a single-star spectrum. Under the joint influence of the micro-turbulence and the line mixing, deriving \logg \space directly from the SB2 spectra underestimates the measured results. Although Traven2020 accounted for the micro-turbulent velocity, the spectral line mixing still makes sense, resulting in the \logg \space measurements that are smaller than the light curve solution.

%__________________________________________________________________

\section{Summary}
\label{sect:summary}

We provide a method to model the binary star spectra with double-lined features. The antecedent information of a detached eclipsing binary system is derived by solving the light curves and radial velocity curves. We synthesise the SB2 model spectra by superposing single star spectra according to the luminosity contribution of the component stars. The synthetic spectra are then used to fit the observation spectra by the least square method and derive atmospheric parameters of the component stars. The method provides radial velocities, effective temperatures and surface gravity of each star, and metallicity for the binary system. We generate model spectra for different phases according to the change of the relative position of the component stars. The atmospheric parameters derived from the multi-epoch SB2 spectra have similar values, which indicates the robustness of our method. 

Our method gives the way to calculate effective projection areas of component stars in both eclipsing and non-eclipsing phases. But in the fitting experiments shown in Sect~\ref{sect:parameters}, we find that the LAMOST MRS SB2 spectra were observed mostly in the non-eclipsing phases. With the resolution of 7500, LAMOST MRS can not resolve double-lined features in nearly eclipsing phases. So the LAMOST MRS SB2 spectra contain fluxes from the whole system of stars.

This method can also be applied to partial eclipsing spectra as long as the resolution is high enough to resolve binary features. Besides, with the increase of resolution, the influence of the limb darkening effect, the reflection between two stars and the microturbulence on the spectral line begins to appear and should be considered in determining atmospheric parameters.

%__________________________________________________________________

\begin{acknowledgements}

The authors thank the reviewer for useful comments to the manuscript. This work is supported by National Science Foundation of China (Nos U1931209, 11970360, 12003050, 12090044, 12103068) and National Key R\&D Program of China(Nos. 2019YFA0405102, 2019YFA0405502).

Guoshoujing Telescope (the Large Sky Area Multi-Object Fibre Spectroscopic Telescope, LAMOST) is a National Major Scientific Project built by the Chinese Academy of Sciences. Funding for the project has been provided by the National Development and Reform Commission. LAMOST is operated and managed by the National Astronomical Observatories, Chinese Academy of Sciences.

This paper includes data collected by the Kepler mission and obtained from the MAST data archive at the Space Telescope Science Institute (STScI). Funding for the Kepler mission is provided by the NASA Science Mission Directorate. STScI is operated by the Association of Universities for Research in Astronomy, Inc., under NASA contract NAS 5–26555.

This paper includes data collected with the TESS mission, obtained from the MAST data archive at the Space Telescope Science Institute (STScI). Funding for the TESS mission is provided by the NASA Explorer Program. STScI is operated by the Association of Universities for Research in Astronomy, Inc., under NASA contract NAS 5–26555.

\end{acknowledgements}

%__________________________________________________________________
\bibliographystyle{aa} % style aa.bst
\bibliography{ref.bib} % your references Yourfile.bib

%\newpage
%\onecolumn
%\newgeometry{margin=5mm,top=5mm}
%begin{landscape}
%\begin{sidewaystable*}
%\begin{flushleft}

%__________________________________________________________________

\begin{appendix}

\section{Projection Areas Calculation}
\label{appendix-a}

1. Basic Assumptions: Spherical component stars.

2. Initial data:

2.1 Fit RV curves and mass ratio q using spectra radial velocities.

2.2 Period(P), eccentricity(e), inclination(i), argument of periastron ($\omega$), time of periastron passage $t_{per}$, semi major axis(SMA), radii of component stars (suppose R$_1>$ R$_2$).

\begin{figure}
    \centering
    \includegraphics[width=\columnwidth]{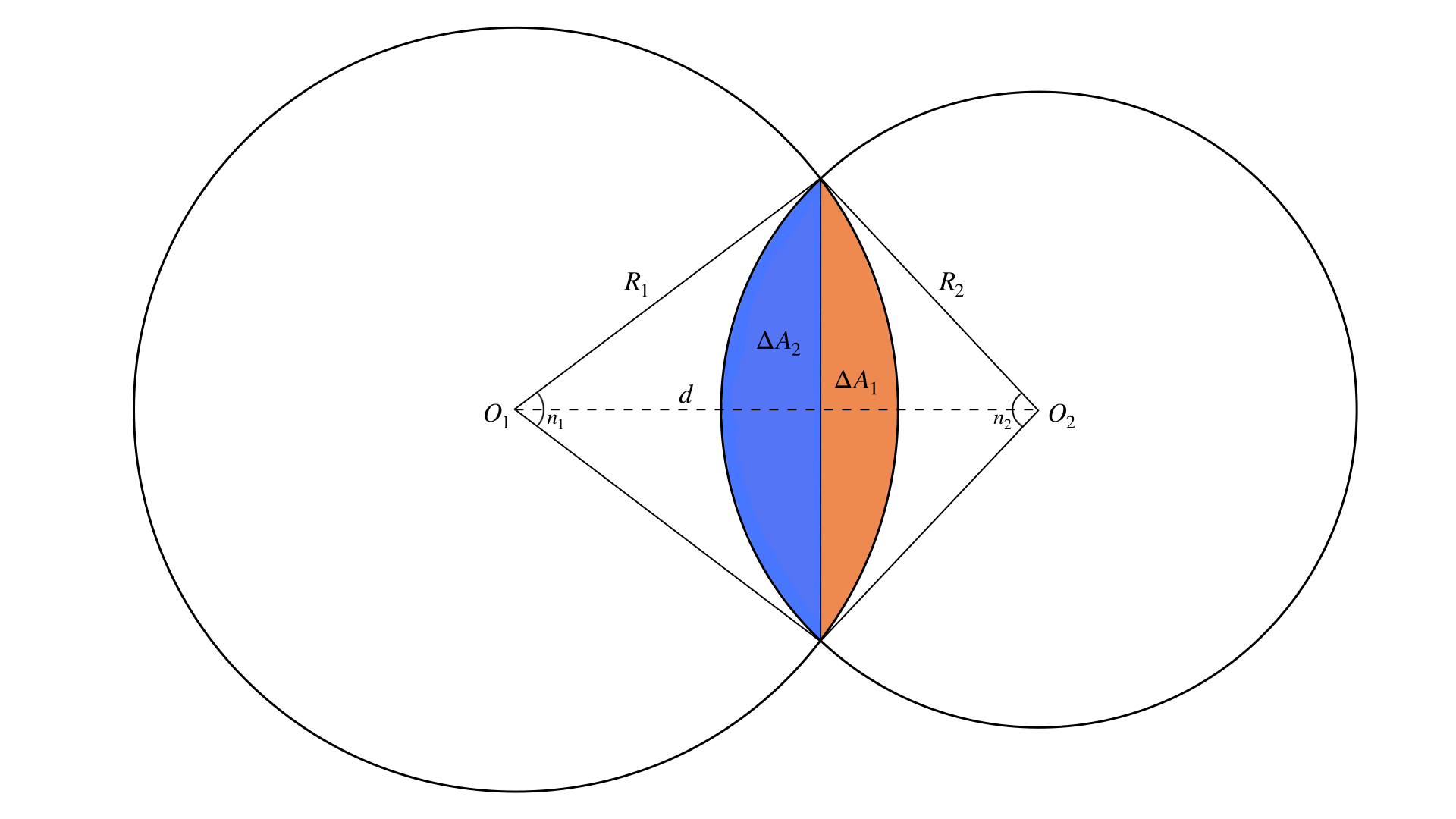}
    \caption{Schematic diagram of the projection areas. $O_1$ and $O_2$ represent the projected centres of the stellar discs, and $d$ is the distance between them. $R_1$ and $R_2$ stand for the radii of the component stars. The areas $\Delta A_1$ and $\Delta A_2$ vary in different orbital phases. See step~3 in Appendix~\ref{appendix-a} for their calculation.}
    \label{fig:app-area}
\end{figure}

3. Projection areas:

Figure~\ref{fig:app-area} is a schematic of the projection areas during shallow partial eclipse. The specific calculation is as follows:

3.1 At the observation time $t$ of an SB2 spectrum, the mean anomaly \it M\rm, the eccentric anomaly \it E \rm and the true anomaly $\upsilon$ of the binary system are calculated as following:
\begin{equation}
		 M = 2 \pi \frac{t-t_{per}}{P},
		\label{equ-app-M}
\end{equation}
\begin{equation}
    E - e \sin E = M,
    \label{equ-app-E}
\end{equation}
\begin{equation}
    \tan \frac{E}{2} = \sqrt{\frac{1-e}{1+e}} \tan \frac{\upsilon}{2}.
    \label{equ-app-upsilon}
\end{equation}

3.2 Projection distance $d$ between the component stars onto the plane-of-sky:
\begin{equation}
		 d = SMA {\frac{1-e^2}{1+e\cos \upsilon}} [1-\sin^2(\omega+\upsilon)\sin^2 i]^{1/2}.
		\label{equ-app-d}
\end{equation}

3.3 Projection areas during different phases:

3.3.1 No eclipse ($d>R_1+R_2$): 
\begin{gather}
        A_{1ini} = \pi \cdot R_1^2, \notag \\
        A_{2ini} =\pi \cdot R_2^2.
\end{gather}

3.3.2 Total and annular eclipse ($d<R_1-R_2$): when the larger one is in front (total eclipse):
\begin{gather}
        A_1 = A_{1ini}, \notag \\
        A_2=0.
\end{gather}

 When the smaller one is in front (annular eclipse):
\begin{gather}
        A_1 = A_{1ini} - A_{2ini}, \notag \\
        A_2 = A_{2ini}.
\end{gather}

3.3.3 Partial eclipse:

a. Shallow partial eclipse ($d>\sqrt{R_1^2 - R_2^2}$):

- larger one in front: 
\begin{gather}
        A_1 = A_{1ini}, \notag \\
        A_2 = A_{2ini} - \Delta A_1 - \Delta A_2.
\end{gather}

- smaller one in front: 
\begin{gather}
        A_1 = A_{1ini} - \Delta A_1 - \Delta A_2, \notag \\
        A_2 = A_{2ini}.
\end{gather}

b. Deep partial eclipse ($\sqrt{R_1^2 - R_2^2}>d>R_1-R_2$): 

- larger one in front: 
\begin{gather}
        A_1 = A_{1ini}, \notag \\
        A_2 = \Delta A_2 - \Delta A_1.
\end{gather}

- smaller one in front: 
\begin{gather}
        A_1 = A_{1ini} - A_{2ini} - \Delta A_1 + \Delta A_2, \notag \\
        A_2 = A_{2ini}.
\end{gather}

$\Delta A_1$ and $\Delta A_2$ are approximated to the area of a portion of a circle cut by a line segment. 
\begin{equation}
    \Delta A_i = \frac{1}2R_i^2(n_i-\sin(n_i)), \ i=1 \ or \ 2.
\end{equation}

$n_1$ and $n_2$ (in radian) in the above equations represent the central angle corresponding to the two circles meeting chord on each projection of the stellar surface. They can be calculated by:
\begin{gather}
        n_1 = 2 \arccos(\frac{d^2+R_1^2-R_2^2}{2dR_1}), \notag \\
        n_2 = 2 \arccos(\frac{d^2-R_1^2+R_2^2}{2dR_2}).
\end{gather}

%\end{tabular}
%\end{longtable}

%\end{tiny}

	\section{Fitting Spectra from Near-Eclipsing Phases}
\label{appendix-b}

\begin{figure*}
	\center
	\includegraphics[width=16.0cm, angle=0]{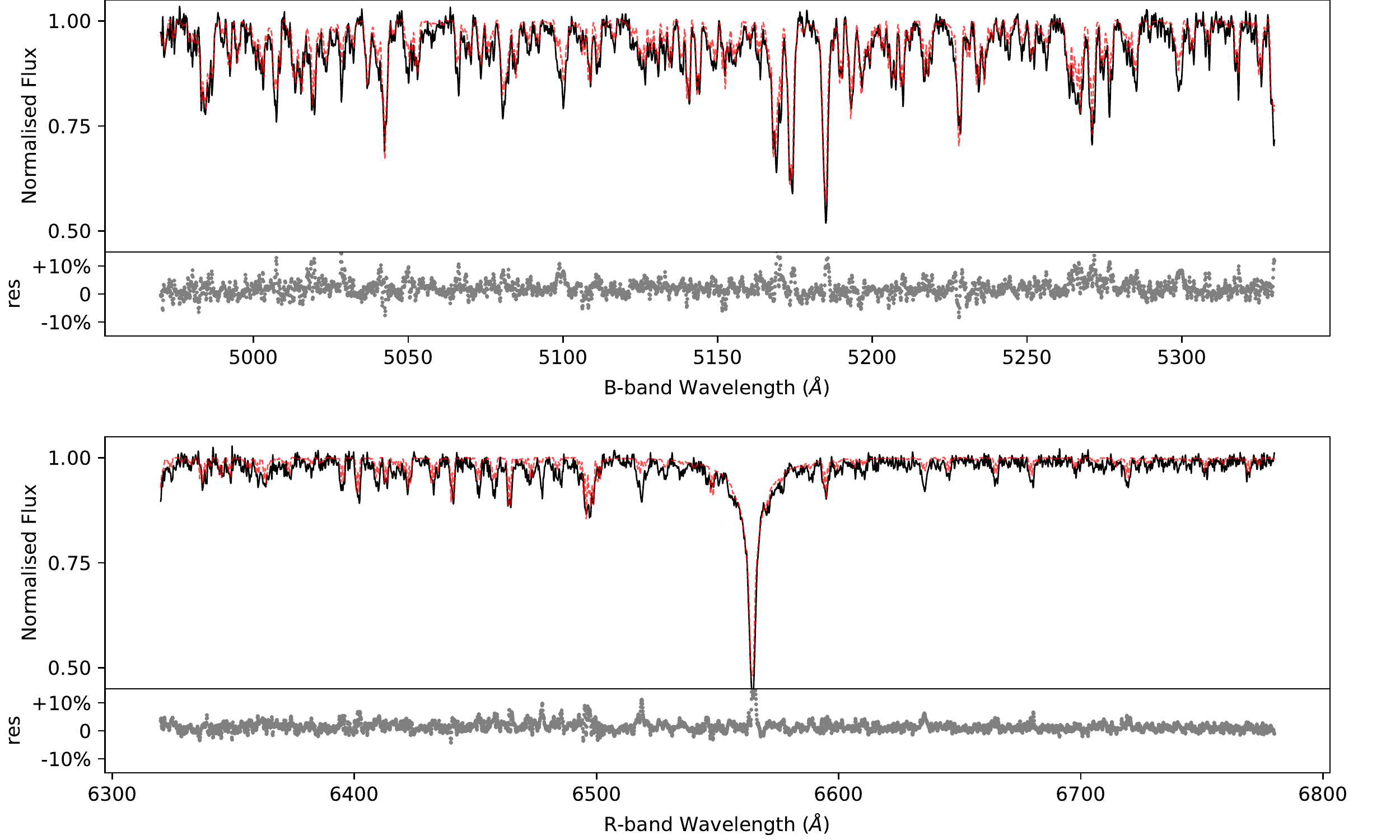}
	\caption{The best-fitting result of TIC 63209649 at the near-eclipsing phase. Lines and colours have the same meaning as in Figure~\ref{eg02fitting}. This spectrum is also the fifth from the top in Figure~\ref{eg02spec}. The orbital phase of this spectrum is 0.9931.}
	\label{ap-eg01}
\end{figure*}

At the near-eclipsing phases, although the double-line features can not be recognised in the LAMOST MRS resolution, the asymmetry of lines is caused when $\Delta RV < 50$ \kms. We present the spectral fitting results of two near-eclipsing spectra. Figure~\ref{ap-eg01} shows the best fitting model of a TIC 63209649 spectrum. The black lines are the LAMOST MRS spectra, the red dashed lines are the best fitting model, and the grey dotted lines represent the residuals. Although the spectrum can be fitted by a synthetic binary spectrum with the help of the orbital results, the atmospheric parameters are quite different from the results of the SB2 spectra. The best fitting parameters of Figure~\ref{ap-eg01} are: $T_{\rm{eff1}}=6263.63 $ K, $\log \, g_{\rm 1}=3.75 $ dex for the primary star, $T_{\rm{eff2}}=6220.43 $ K, $\log \, g_{\rm 2}=4.44 $ dex for the secondary star, and $ \rm[M/H]= -0.48 $ dex for the binary system. But the mean parameters of the SB2 spectral fitting are: $T_{\rm{eff1}}=6643.22 \pm 49.20 $ K, $\log \, g_{\rm 1}=4.13 \pm 0.13 $ dex for the primary star, $T_{\rm{eff2}}=5917.39 \pm 37.60 $ K, $\log \, g_{\rm 2}=3.78 \pm 0.04 $ dex for the secondary star, and $ \rm [M/H]= -0.23 \pm 0.13 $ dex for the binary system (Table~\ref{tab-eg02specparam}). In Sect.~\ref{subsect:conti}, we find the atmospheric parameters derived in different phases should not vary a lot if the contributions of the component stars are detected correctly. Therefore, the fitting results in the figure are likely to be inconsistent with the real situation. We suggest that the real signal-to-noise ratio (S/N) of the near-eclipsing spectrum is not high enough to recognise the line asymmetry correctly, despite the announced S/N of this spectrum being 60.42 for the blue band and 94.05 for the red band in the LAMOST MRS catalogue.

\begin{figure*}
	\center
	\includegraphics[width=16.0cm, angle=0]{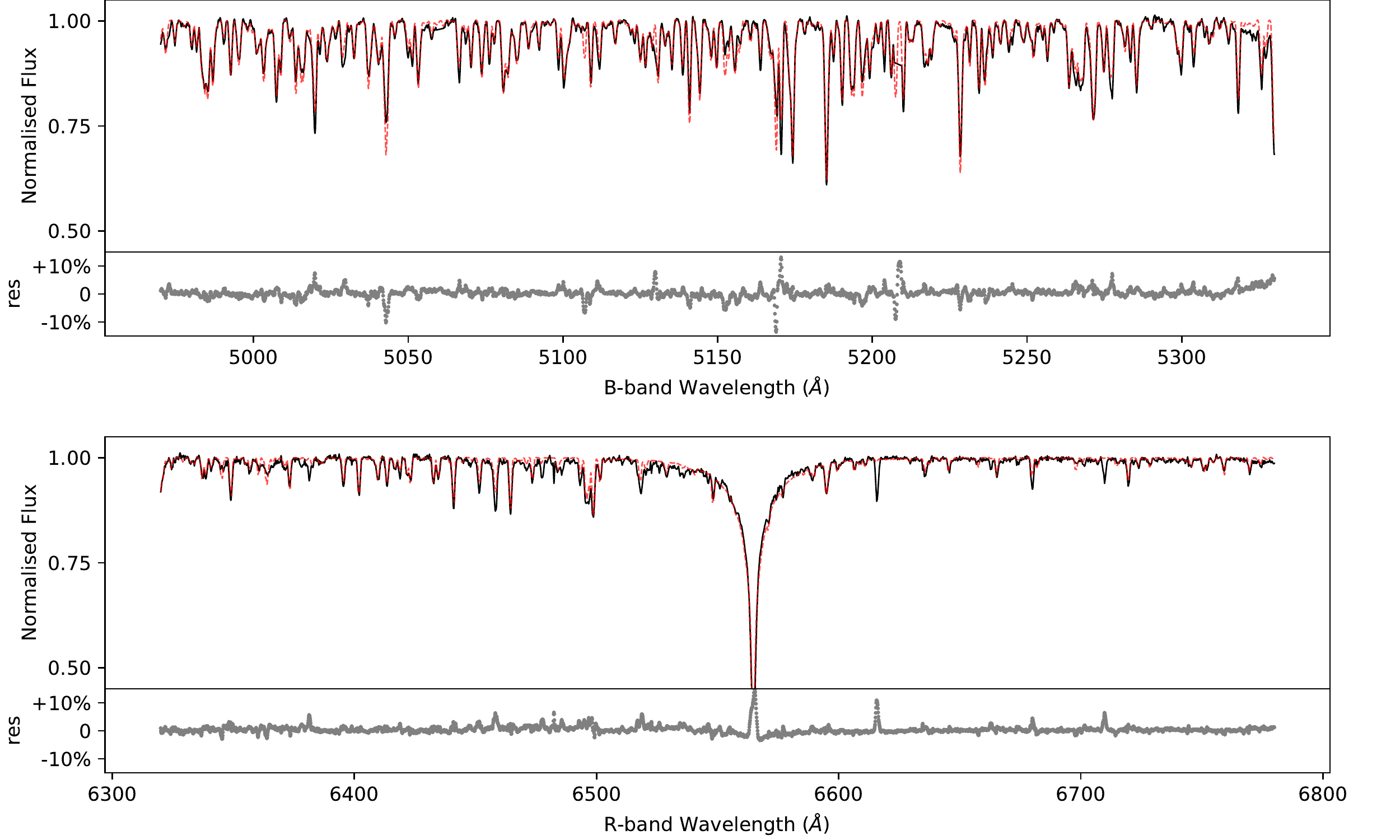}
	\caption{The best-fitting result of EPIC 247529791 at the near-eclipsing phase. Lines and colours have the same meaning as in Figure~\ref{ap-eg01}. This spectrum is also the fourth from the top in Figure~\ref{eg01spec}. The orbital phase of this spectrum is 0.0260.}
	\label{ap-eg02}
\end{figure*}

Figure~\ref{ap-eg02} is the near-eclipsing phase fitting result of EPIC 247529791. The atmospheric parameters fitted from these spectra are different from the parameters from the SB2 spectra, too. The best fitting parameters of Figure~\ref{ap-eg02} are: $T_{\rm{eff1}}=6905.15 $ K, $ \log \, g_{\rm 1}=3.60 $ dex for the primary star, $T_{\rm{eff2}}=6367.69 $ K, $\log \, g_{\rm 2}=3.32 $ dex for the secondary star, and $\rm [M/H]= -0.31 $ dex for the binary system. But the mean parameters of the SB2 spectral fitting are: $T_{\rm{eff1}}=6969.73 \pm 16.30 $ K, $\log \, g_{\rm 1}=3.56 \pm 0.06 $ dex  for the primary star, $T_{\rm{eff2}}=6146.84 \pm 38.50 $ K, $\log \, g_{\rm 2}=3.39 \pm 0.30 $ dex for the secondary star, and $\rm [M/H]= -0.46 \pm 0.02 $ dex for the binary system (Table~\ref{tab-eg01specparam}). We don't use the parameters of the near-eclipsing spectra to calculate the final parameters in this work, but we note that our method has the ability to deal with the line-asymmetry spectra with more precise orbital parameters and high S/N spectra.

\end{appendix}

%\end{flushleft}
%\end{landscape}
%\begin{sidewaystable*}
%\newgeometry{margin=1.3cm,top=2cm,bottom=2.5cm}

\twocolumn
%}
\end{document}